\documentclass[authoryear,preprint,review,12pt]{elsarticle}
\pdfoutput=1 
\usepackage{latexsym}
\usepackage{natbib}
\usepackage{amsmath}%permite usar las abreviaciones de los simbolos como dfrac
\usepackage{amsfonts}%permite escribir con los simbolos matematicos, las fuentes de la american mathematical society
\usepackage{amssymb}%permite declarar nuevos comandos para simbolos: \DeclareMathSymbol
\usepackage[T1]{fontenc}
\usepackage{pdfpages}
\usepackage{epstopdf}
\usepackage{graphicx}  
\usepackage{verbatim}   % useful for program listings
\usepackage{t1enc}
\usepackage{anysize}
\usepackage{xcolor,soul}
\usepackage{tikz}
\usepackage{subfig}
\usepackage[percent]{overpic} %esto es para enumerar a las subfiguras: (a),(b),etc.
\usepackage{float}  %esto es para ubicar las figuras donde yo quiero. Hay que poner [H] y no [h] como hacía siempre.
\usepackage{appendix} %sirve para poner apéndice
\usepackage{booktabs} %booktabs es para hacer tablas pro.
\usepackage{natbib}
\usepackage{multirow}%es para poder unir filas en una tabla
\usepackage{array}
\usepackage{chngcntr}

\DeclareGraphicsExtensions{.pdf,.eps}

\begin{document}

\begin{frontmatter}

\title{Effects of rainfall on {\it Culex} mosquito population dynamics}
\author[label1]{L.D. Valdez}
\author[label1]{G.J. Sibona}
\author[label2,label3]{L.A. Diaz} 
\author[label3]{M.S. Contigiani}
\author[label1]{C.A. Condat}

 \address[label1]{Facultad de Matem\'atica, Astronom\'ia, F\'isica y Computaci\'on, Universidad Nacional de C\'ordoba, Instituto de F\'isica Enrique Gaviola, CONICET, Ciudad Universitaria, 5000 C\'ordoba, Argentina}
\address[label2]{Instituto de Investigaciones Biol\'ogicas y Tecnol\'ogicas--CONICET--Universidad Nacional de C\'ordoba, C\'ordoba, Argentina}
\address[label3]{Laboratorio de Arbovirus--Instituto de Virolog\'ia ``Dr. J. M. Vanella''--Facultad de Ciencias M\'edicas--Universidad Nacional de C\'ordoba, C\'ordoba, Argentina}

%%\fntext[fn1]{This is the specimen author footnote.}
\begin{abstract}
  The dynamics of a mosquito population depends heavily on climatic
  variables such as temperature and precipitation. Since climate
  change models predict that global warming will impact on the
  frequency and intensity of rainfall, it is important to understand
  how these variables affect the mosquito populations. We present a
  model of the dynamics of a {\it Culex quinquefasciatus} mosquito
  population that incorporates the effect of rainfall and use it to
  study the influence of the number of rainy days and the mean monthly
  precipitation on the maximum yearly abundance of mosquitoes
  $M_{max}$. Additionally, using a fracturing process, we investigate
  the influence of the variability in daily rainfall on $M_{max}$. We
  find that, given a constant value of monthly precipitation, there
  is an optimum number of rainy days for which $M_{max}$ is a
  maximum. On the other hand, we show that increasing daily rainfall
  variability reduces the dependence of $M_{max}$ on the number of
  rainy days, leading also to a higher abundance of mosquitoes for the
  case of low mean monthly precipitation. Finally, we explore the
  effect of the rainfall in the months preceding the wettest season,
  and we obtain that a regimen with high precipitations throughout the
  year and a higher variability tends to advance slightly the time at
  which the peak mosquito abundance occurs, but could significantly
  change the total mosquito abundance in a year.
\end{abstract}

\end{frontmatter}

\section{Introduction}
Mosquito-transmitted flaviviruses are an increasing health threat. In
particular, members of {\it Culex} species (Diptera: Culicidae) such as
{\it Cx. quinquefasciatus} and {\it Cx. interfor} are responsible for
transmitting the West Nile and St. Louis encephalitis (SLEV)
viruses to humans and domestic
animals~(\cite{beltran2015evidencia,diaz2008eared,diaz2016activity,gubler2002global,gubler2007flaviviruses,lumsden1958st}). For instance, the SLEV is endemic in Argentina,
where the principal vector is postulated to be {\it
  Cx. quinquefasciatus}~(\cite{diaz2013transmission}). In the last
decades, mosquito-borne diseases have emerged and re-emerged as a
result of multiple factors such as increasing urbanization,
international travel, and climate
change~(\cite{harrigan2014continental,kilpatrick2011globalization}). The
development of mathematical models is essential to quantify the effect
of each of these factors on the dynamics of the mosquito population,
and to determine the most effective strategies to control the epidemic
outbreaks transmitted by mosquito
vectors~(\cite{And_01,ewing2016modelling,lord2001simulation,lord2001simulation0,marini2016role}).

Multiple studies have shown that the life cycle of the {\it
  Cx. quinquefasciatus} is closely related to
temperature~(\cite{almiron1996winter,ciota2014effect,gunay2011narrow,loetti2011development,strickman1988rate}). \cite{strickman1988rate}
demonstrated that its reproductive activity increases with
temperature, and~\cite{almiron1996winter} showed that this species can
only live in environments with a temperature above
10$^{\circ}$C. Other studies have found that temperature has a strong
influence on the development and survival of both adult and immature
mosquitoes~(\cite{ciota2014effect,gunay2011narrow,loetti2011development}). In turn, it was observed that {\it
  Cx. quinquefasciatus} does not enter diapause, but it may undergo
quiescence or remain gonoactive in protected (indoor or underground)
habitats~(\cite{almiron1996winter,nelms2013overwintering}). Urbanization
therefore helps {\it quinquefasciatus} populations survive the mild
winters of temperate regions.

Similarly, rainfall is an important climatological variable to predict
the abundance of {\it Culex} mosquitoes, since its copiousness and
distribution determine the production and size of mosquito breeding
sites. \cite{reisen2008impact} studied the changes in the {\it Cx.
  tarsalis} population in California and found that, in most regions,
it is positively correlated with an increase in total
precipitation. However, these authors also found that in some places of
the driest region of California, the correlation between these two
variables was negative. On the other hand,~\cite{olson1983correlation}
showed that a very large rainfall is not always accompanied by
proportionately large increases in the abundance of {\it
  Cx. tritaeniorhynchus} and {\it Cx. gelidus}. In consequence, these
results indicate that there exists a nonlinear relationship between
rainfall and {\it Culex} abundance, which should be modeled in order
to predict mosquito abundance. Additionally, since the climatological
projections suggest that global warming will alter the frequency and
intensity of rainfall, it is crucial to understand how different
rainfall patterns will affect mosquito populations.

In this paper we develop a dynamic model of the {\it
  Cx. quinquefasciatus} population, adapting a fracturing
procedure~(\cite{finley2014exploring}) to describe the rainfall
distribution. We use a system of compartmental ordinary differential
equations that describe the immature and adult mosquito populations, in
which we introduce the influence of temperature and rainfall on the
reproduction rate. To study the influence of different rainfall
patterns, we use a synthetic time series of rainfall based on the
amount of rainfall per month and the monthly number of rainy days.  We
find that, for a given constant value of monthly precipitation, there
is an optimum number of rainy days for which the maximum $M_{max}$ in
the mosquito population is highest.

On the other hand, we also study the variability of daily rainfall
intensity through a fracturing process~(\cite{finley2014exploring}),
which allows us to study homogeneous and heterogeneous rainfall
regimes, including those characterized by heavy rain events. We show
that increasing daily rainfall variability reduces the dependence of
$M_{max}$ on the number of rainy days, leading also to a higher
abundance of mosquitoes for the case of low mean monthly
precipitation.

Finally, we explore the effect of different winter precipitation
regimes on the mosquito abundance in the summer season, obtaining that
a higher variability tends to advance slightly the peak time of
mosquito abundance. Interestingly, we predict that the accumulated
abundance of mosquitoes will decrease in a regime with high
variability in the rainfall intensity.

The boundary of the {\it Cx. quinquefasciatus} habitat in South
America runs across central Argentina. This region is thus expected to
exhibit intense changes in mosquito populations due to the undergoing
climatic change; this is likely to have a strong impact on flavivirus
prevalence. For this reason, we use the climatic data for the city of
C\'ordoba to calibrate our model in the period 2008-2009.

The paper is organized as follows: in Sec.~\ref{Sec.modd} we
present the model of the dynamics of mosquito population and in
Secs.\ref{SecConst} and \ref{SecVarRai} we explain the methods for
generating two different synthetic time series. Then in
Sec.~\ref{secResul} we show our results and finally we present our
conclusions in Sec.~\ref{sec.Discussion}.

\section{Methods}
\subsection{The model of mosquito abundance}\label{Sec.modd}

In this section, we construct a compartmental, ordinary differential
equation model for the mosquito abundance. We consider that the total
vector population is stage-structured with an immature class
consisting of all aquatic stages, and a mature or adult class . We
assume that these population groups are restricted only to female
mosquitoes as the reproductive sex.

In our model, the total birth rate, $i.e.$ the total number of new
immature female mosquitoes per unit of time, is proportional to the
number of adult mosquito females and to $\beta_L\lambda(t) \theta(t)$,
where $\beta_L$ corresponds to the reproduction rate in optimal
conditions of temperature and water availability, and $\lambda(t)$ and
$\theta(t)$ are normalized factors describing the influence of
rainfall and temperature on the total birth rate, respectively. In
Sec.~\ref{Sec.lambThe} we will explain how we construct these
factors. In addition, we assume that the total birth rate is also
regulated by a carrying capacity effect that depends on the occupation
of the available immature habitats. Therefore we propose that the
immature population growth is logistic-like with a carrying capacity
$K_L$. Additionally, immature individuals either go to the mature
class with rate $m_L$ or die at a rate $\mu_L$. We stress that the
ratio $1/m_L$ gives the average development time from immature
mosquito to adult. Finally, adult mosquitoes die with a mortality rate
$\mu_M$. For simplicity, we assume that $m_L$, $\mu_L$ and $\mu_M$ do
not depend on temperature or rainfall. With these definitions, we
propose the following dynamic mass-balance equations for the abundance
of immature mosquitoes, $L(t)$, and adult mosquitoes, $M(t)$,

\begin{eqnarray}
  L(t+\Delta t)&=& L(t)+\Delta t \left[ \beta_L \theta(t) \lambda(t) M(t)\left(1-\frac{L(t)}{K_L}\right)-m_L L(t)-\mu_L L(t)\right],\label{eq.larv}
\end{eqnarray}
and
\begin{eqnarray}
M(t+\Delta t)&=& M(t)+\Delta t \left[ m_L L(t)-\mu_M M(t)\right],\label{eq.adult}
\end{eqnarray}
where $\Delta t$ is the time step size. Here we use $\Delta t=0.1$
[days]. Equation~(\ref{eq.larv}) can be easily derived by assuming
that the fraction of the carrying capacity corresponding to female
immature mosquitoes is the same as the fraction of females in the
immature population. Table~\ref{tab.Trans} summarizes the different
state variables and parameters used in this paper.
 
\begin{table}[H]
\centering
\caption{The variables and parameters for Eqs.~(\ref{eq.larv})-(\ref{eq.ArtD}).}
\label{tab.Trans}
\begin{tabular}{|c|p{9.5cm}|c|c|}
\hline
Quantity & Definition & Value & Refs. \\
\hline
$L$ &  number of immature female mosquitoes &---&\\
$M$&  number of adult female mosquitoes &---&\\
$\beta_L$& birth rate of immature female mosquito per female adult mosquito in optimal conditions of temperature and water availability (days$^{-1}$) &13.5&fitted (see \ref{AppCalib}) \\
$\theta$ & effect of the temperature on the birth rate of mosquitoes  &---&\\
$\lambda$& effect of the water availability on the birth rate of mosquitoes   &---&\\
$m_L$&  rate at which immature mosquitoes develop into adults (days$^{-1}$)  &0.098&\cite{loetti2011development}\\
$\mu_L$& immature mortality rate (days$^{-1}$)  &0.03&\cite{loetti2011development}\\
$\mu_M$& mosquito mortality rate (days$^{-1}$)  &0.078& \cite{david2012bionomics}\\
$K_L$& carrying capacity of the immature female population &16.2&fitted (see \ref{AppCalib})\\
$H_{max}$&maximum daily amount of accumulated rainwater [mm]&9.86&fitted (see \ref{AppCalib})\\
$H_{min}$&minimum daily amount of accumulated rainwater [mm]&0.067&fitted (see \ref{AppCalib})\\
$H$&accumulated amount of rainwater [mm]&---&\\
$R$&daily rainfall [mm]&---&\\
$E$&daily evapotranspiration [mm]&---&\\
$T$&average daily temperature [$^{\circ}$C]&---&\\
$Hum$&daily relative humidity&---&\\
$P_{max}$&total rainfall in the wettest month [mm]&---&\\
$P_{min}$&total rainfall in the driest month [mm]&---&\\
$D_{max}$&total number of rainy days in the wettest month [days]&---&\\
$D_{min}$&total number of rainy days in the driest month [days]&---&\\
\hline
\end{tabular}
\end{table}

\subsubsection{Effect of temperature and rainfall on the total birth rate}\label{Sec.lambThe}

We add the effect of the temperature through a temperature factor
$\theta(t)$. Several studies have shown that {\it Cx. quinquefasciatus} can breed only
at temperatures above
10$^{\circ}$C~(\cite{almiron1996winter,ribeiro2004thermal}) and that
the number of egg rafts collected per day is closely correlated with
temperature~(\cite{strickman1988rate}). Therefore we propose that the
factor $\theta$(t) is a piecewise linear function,

\medskip
%\[\nonumber
\begin{eqnarray}\label{cpattern}
\theta(t)=\left\{%
\begin{array}{ll}
\frac{T(t)-T_{A}}{T_{B}-T_{A}} &\;\;\;\; \text{if}\;\;\;\; T_{A}\leqslant T(t) \leqslant T_{B} \\
1  & \;\;\;\; \text{if}\;\;\;\;T(t)>T_{B}\\
0  &\;\;\;\; \text{if}\;\;\;\; T(t)<T_{A},
\end{array}%
\right.
\end{eqnarray}

%\]
\medskip where $T(t)$ is the average daily temperature and
$T_{A}=10^{\circ}$C corresponds to a minimum temperature below which
the net birth rate vanishes. The choice of the function $\theta(t)$ is
not unique: for instance, a power law may be used instead
(\cite{ewing2016modelling}). Here we assume that the positive effect
of the temperature on the birth rate saturates at $T_B$, since it was
observed by~\cite{oda1980effects} that the number of egg rafts per
female does not significantly change between 21$^{\circ}$C and
30$^{\circ}$C.

On the other hand, mosquito reproduction is also triggered by
rainfalls since these increase the number of breeding sites, such as
temporary ground pools.  In order to introduce the effect of
rainfall on the mosquito birth rate, we compute the accumulated amount
of rainwater $H$, whose variation is given
by the total daily rainfall $R(t)$ minus the evapotranspiration $E(t)$~(\cite{gong2011climate}),
\begin{eqnarray}\label{eq.HRE}
H(t+1)=H(t)+ \left[ R(t)-E(t) \right].
\end{eqnarray}
The function $H(t)$ is a convenient measure of the quantity of water
available for breeding sites. It should represent the average level of
puddles, ponds, drains, small streams, and underground sources such as
waste water channels in urban environments. As we will explain below,
equation~(\ref{eq.HRE}) is applied on a daily time scale. The term of
evapotranspiration is estimated using the Ivanov
model~(\cite{romanenko1961computation, valipour2014application}), which
is based on the mean temperature and the relative humidity [$Hum(t)$],
and it is given by
\begin{eqnarray}\label{eq.EvapEq}
E(t)=6.10^{-5}(25+T(t))^2(100-Hum(t)).
\end{eqnarray}
Note that the evapotranspiration $E(t)$ is a monotonically increasing
function of temperature and a decreasing function of humidity. In
particular, $E(t)$ vanishes for $Hum=100$\%. In addition, we assume
that the level of accumulated water $H$ varies only between a minimum
($H_{min}$) and a maximum ($H_{max}$) boundary level, $i.e.$,

\medskip
%\[\nonumber
\begin{eqnarray}\label{cpattern}
H(t+1)=\left\{%
\begin{array}{ll}
H_{min} &\;\;\;\; \text{if}\;\;\;\; H(t)+ \left[ R(t)-E(t) \right]\leqslant H_{min} \\
H_{max}  & \;\;\;\; \text{if}\;\;\;\;H(t)+\left[ R(t)-E(t) \right]\geqslant H_{max}\\
H(t)+ \left[ R(t)-E(t) \right] &\;\;\;\; \text{otherwise}. \\
\end{array}%
\right.
\end{eqnarray}

%\]
\medskip
Here, $H_{min}$ represents a minimum amount of water that is always
available for mosquito breeding, for instance in permanent streams or
in the drainage system, while $H_{max}$ is a level of water above
which the breeding sites overflow~(\cite{karl2014spatial}).

Finally, the factor $\lambda(t)$ that takes into account the effect of
 rainfall on the birth rate (see Eq.~(\ref{eq.larv})), is the
normalization of $H(t)$:
\begin{equation}\label{eq.lamFin}
\lambda(t)=\frac{H(t)}{H_{max}}.
\end{equation}
Note that the minimum value of $\lambda(t)$ is $H_{min}/H_{max}$.

Although we take the integration time step to be $\Delta t=0.1$ days,
the data of temperature, rainfall and humidity are available only
on a daily time scale. Therefore, for all the integration time steps
within a day ``$d$'' (i.e. $d\leq t<d+1$), we set $\lambda(t)$ and
$\theta(t)$ to have the values computed through Eqs.~(\ref{cpattern})
and~(\ref{eq.EvapEq}) using the corresponding meteorological data for
day ``$d$''.

\subsection{Model of synthetic rainfall using the monthly number of rainy days and the monthly precipitation}\label{SecConst}

The amount of average rainfall $P$ and the number of days $D$ with
rain per month are two parameters commonly used to characterize the
long-term precipitation
trend~(\cite{madsen2009update,owusu2013changing,zhai2005trends}). In
this section, we show how they can be used to construct a
synthetic rainfall time series.

The average monthly rainfall is assumed to follow a sinusoidal function,
\begin{eqnarray}\label{eq.ArtP}
P(m)= \frac{P_{max}-P_{min}}{2}\cos\left(\frac{2\pi}{12}(m-m_0)\right)+\frac{P_{max}+P_{min}}{2},
\end{eqnarray}
where $m=1,...,12$ represents the month (with $m=1$ for January and
$m=12$ for December), $m_0$ corresponds to the month of maximum
rainfall, $P_{max}$ is the total precipitation of the wettest month
$m_0$, and $P_{min}$ is the total precipitation of the driest month,
corresponding to $m = m_0 + 6$. Note that for a higher value of either
$P_{max}$ or $P_{min}$ there is an increase in the annual amount of
precipitation but, while a higher $P_{max}$ enhances the precipitation
difference between the rainy and dry seasons, a higher $P_{min}$
reduces this difference.

\begin{figure}[H]
\centering
\vspace{0.5cm}
\begin{overpic}[scale=0.30]{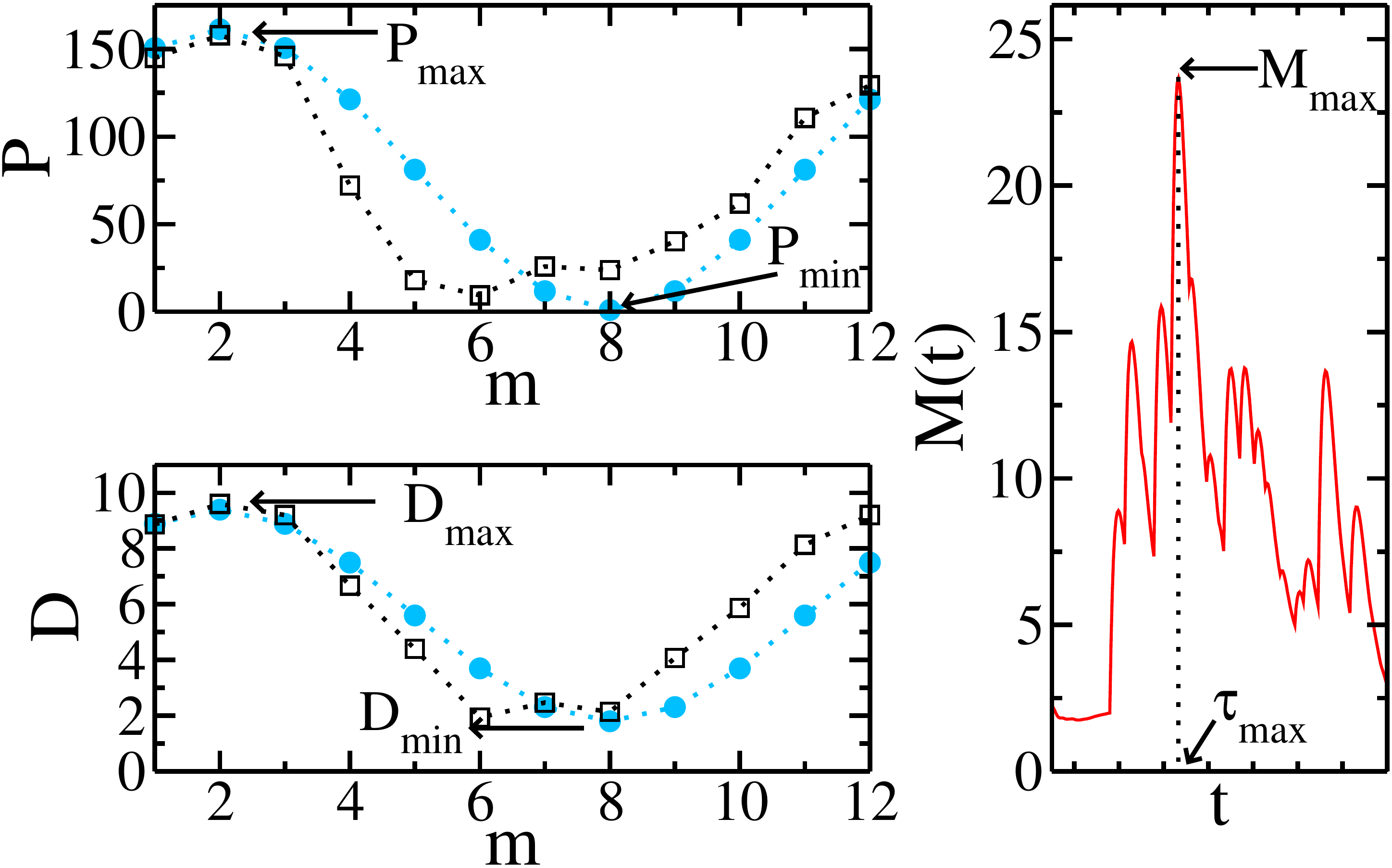}
  \put(40,55){(a)}
  \put(40,20){(b)}
  \put(76,55){(c)}
  \end{overpic}\hspace{0.5cm}\vspace{1.0cm}
\caption{Schematic representation of Eq.~(\ref{eq.ArtP}) (panel (a))
  and Eq.~(\ref{eq.ArtD}) (panel (b)), where $P(m)$ is the monthly amount
  of rainfall and $D(m)$ is the number of rainy days per month (blue
  circles). In these plots we also show the average monthly
  precipitation and number of rainy days in the city of C\'ordoba in
  the period 2001-2015 (black squares)~(\cite{Exxon_01}).  Panel (c) is
  a cartoon of $M(t)$ indicating the maximum abundance of mosquitoes
  $M_{max}$ and the time $\tau_{max}$ (in day units) at which this
  peak is reached. Dotted lines are to guide the eye.}\label{fig.Esq1}
\end{figure}

Similarly, we propose that the number of rainy days is given by
\begin{eqnarray}\label{eq.ArtD}
D(m)= \frac{D_{max}-D_{min}}{2}\cos\left(\frac{2\pi}{12}(m-m_0)\right)+\frac{D_{max}+D_{min}}{2},
\end{eqnarray}
where $D_{max}$ and $D_{min}$ correspond to the number of rainy days
in the months labeled by $m_0$ and $m_0 + 6$, respectively.  Choosing
$m_0=2$, this distribution would be suitable for the city of
C\'ordoba. In Fig.~\ref{fig.Esq1}(a) and (b) we show a schematic of
the parameters used in Eqs.~(\ref{eq.ArtP}) and~(\ref{eq.ArtD}).

From Eqs.~(\ref{eq.ArtP}) and~(\ref{eq.ArtD}) we construct the amount
of daily rainfall $R(t)$ (see Eq.~(\ref{eq.HRE})), placing the rainy
days in each month at random and assuming that the amount of rain
specified in Eq.~(\ref{eq.ArtP}) is equally distributed over these
days. Then we calculate the factor $\lambda(t)$ and integrate
Eqs.~(\ref{eq.larv}) and~(\ref{eq.adult}). For simplicity we use in
these equations the values of humidity and temperature obtained from
meteorological data.

It is important to note that, by construction, $R(t)$ is a stochastic
time series since the rainy days for each month are chosen at random;
therefore our results for the effect of the synthetic series $R(t)$ on
the mosquito abundance must be averaged over a large number (we take
$10^4$) of realizations. Using this model of time series of rainfall,
we measure the highest peak of mosquito abundance $M_{max}$ and the
time $\tau_{max}$ at which this peak is reached (see
Fig.~\ref{fig.Esq1}(c)).

In the following section, we explain how to
introduce variability on the amount of daily rainfalls.

\subsection{Model of synthetic rainfalls with variable daily intensity of precipitation}\label{SecVarRai}
In general, the daily rainfall intensity can range from drizzles with
less than $1$~mm to torrential downpours exceeding
200~mm~(\cite{li2013dry,lei2008effect}). Various distributions, such
as Weibull~(\cite{suhaila2007fitting}),
lognormal~(\cite{cho2004comparison}) and generalized
Pareto~(\cite{deidda2010multiple}), have been proposed to model the
precipitation amount. While the above-mentioned distributions could be
used to generate a sequence of rainfalls with variable or
heterogeneous intensity, the disadvantage of this approach is that the
total monthly rainfall is also a stochastic variable. In order to
isolate the effect of the heterogeneity of the rainfalls, we will use
a fracturing or fragmentation process (FT), which allows us to
maintain the total monthly rainfall constant. This method is related
to a cascading procedure that was used by physicists to study the
fragmentation of brittle material~(\cite{hernandez2003two}). Recently
the FT process was also applied to obtain empirical distributions with
a finite tail~(\cite{finley2014exploring}). From a geometrical point
of view~(\cite{borgos2000partitioning,finley2014exploring}), this
process performs a sequential breakage of a segment or interval of
length $\ell$ to obtain $D$ subintervals with variable length
$\widetilde{\ell}$, which can be used to decompose the total monthly
rainfall into daily rainfalls with heterogeneous intensity. In this
representation, $\ell$ and $\widetilde{\ell}$ stand for the total
amount of rainfall in a month and in a day,
respectively. In~\ref{AppFT} we explain the steps of the FT process in
detail.

\begin{figure}[H]
\centering
\vspace{0.5cm}
  \begin{overpic}[scale=0.30]{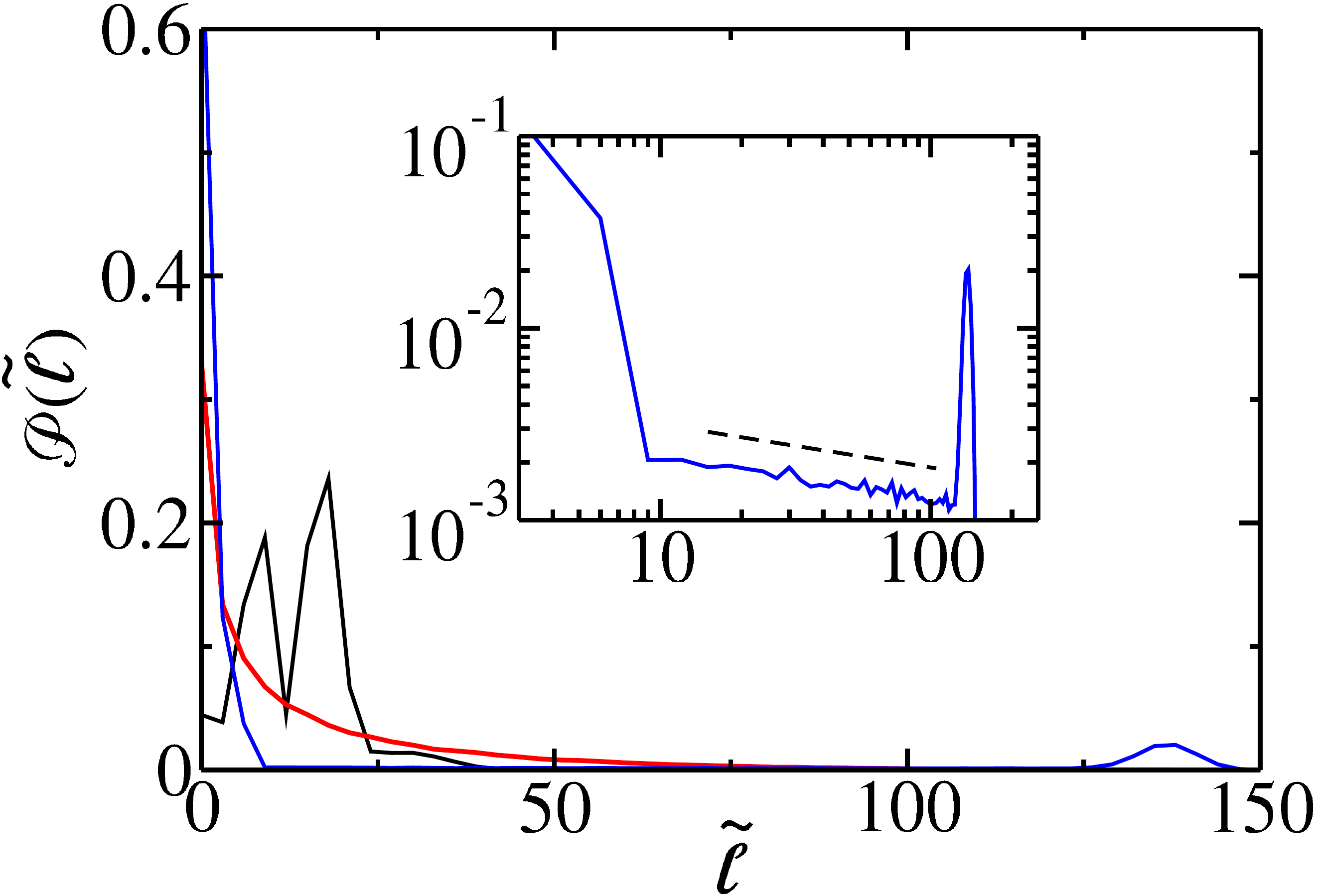}
    \put(85,50){}
  \end{overpic}\hspace{0.5cm}\vspace{0.5cm}
\caption{Distribution of length segments
  $\mathcal{P}(\widetilde{\ell})$, using $\ell=150$~mm (total length) and $D=10$ (number of
  rainy days) for different values of $\alpha$: 0.1 (black), 0.5 (red),
  and 0.9 (blue). These distributions were obtained over $2.10^4$
  realizations. In the inset, we show $\mathcal{P}(\widetilde{\ell})$ for $\alpha=0.9$ in
  log-log scale, and the dashed line corresponds to a power-law fit
  with an exponent 0.22. }\label{fig.alff}
\end{figure}

This method depends on a parameter $\alpha \in [0,1]$ which controls the
heterogeneity of the segment length. In particular:
\begin{itemize}
\item $\alpha=0$ corresponds to the case where an interval is split
  into two subintervals of equal length $\widetilde{\ell}=\ell/2$,
\item $\alpha=1$ corresponds to the special case where two "intervals"
  are generated, one of length zero and the other of length $\widetilde{\ell}=\ell$.
\end{itemize}
In Fig.~\ref{fig.alff} we show how $\alpha$ controls the shape of the
fragment length distribution $\mathcal{P}(\widetilde{\ell})$, using $\ell=150$~mm and
$D=10$. Note that the resulting length of each subinterval
corresponds to the intensity of rainfall in one day.

For small values of $\alpha$ the distribution is concentrated around
the mean value close to $\ell/D$, while for intermediate values of
$\alpha$, the distribution of lengths has a longer tail. Finally, for
high values of $\alpha$, $\mathcal{P}(\widetilde{\ell})$ has a peak near $\ell$ which
depicts a regime where most of the corresponding month total
rainfall is confined to one day. Interestingly, we also note that in
this case the rainfall distribution has a region in which it decays as
a power law.

In order to model the temporal variation of precipitation $R(t)$ (see
Sec.~\ref{Sec.lambThe}), we apply a fracturing process (FT) for each
month, partitioning an interval whose length is the amount of monthly
precipitation given by Eq.~(\ref{eq.ArtP}) and where the number of
subintervals (the number of rainy days) is given by
Eq.~(\ref{eq.ArtD}). See~\ref{AppFT} for further details on
the construction of $R(t)$.

\section{Results}\label{secResul}

\subsection{Calibration}
We calibrate our model of mosquito abundance using a
Metropolis-Hastings algorithm (see~\ref{AppCalib}) with the
number of female {\it Culex quinquefasciatus} mosquitoes collected in C\'ordoba city
($31^\circ$24$'$30$''$ S, $64^\circ$11$'$02$''$ W, C\'ordoba province,
Argentina) every two weeks from January 2008 to December 2009 (see
~\cite{batallanthesis,batallan2015st} for details on the data and their
sources). The climate of C\'ordoba is temperate with dry winters and
hot rainy summers. The mean annual temperature ranges between
$16^\circ$C-$17^\circ$C and the mean annual rainfall is
800~mm~(\cite{jarsun2003caracterizacion}). The temperature [$T(t)$],
relative humidity [$Hum(t)$], and rainfall [$R(t)$] data for C\'ordoba
were obtained from the website~\cite{Exxon_01}.  We
calibrate the following parameters: $\beta_L$, $H_{max}$, $H_{min}$,
and $K_L$. In~\ref{Sec.Sensit} we perform a sensitivity
analysis of these calibrated parameters.

As initial conditions of Eqs.~(\ref{eq.larv}) and~(\ref{eq.adult}), we
set $M(t)=L(t)=20$ . To attenuate the effect of these initial
conditions, the integration of the equations starts 12 months before
we implement the fitting and study our model.

\begin{figure}[H]
\centering
\vspace{0.5cm}
  \begin{overpic}[scale=0.25]{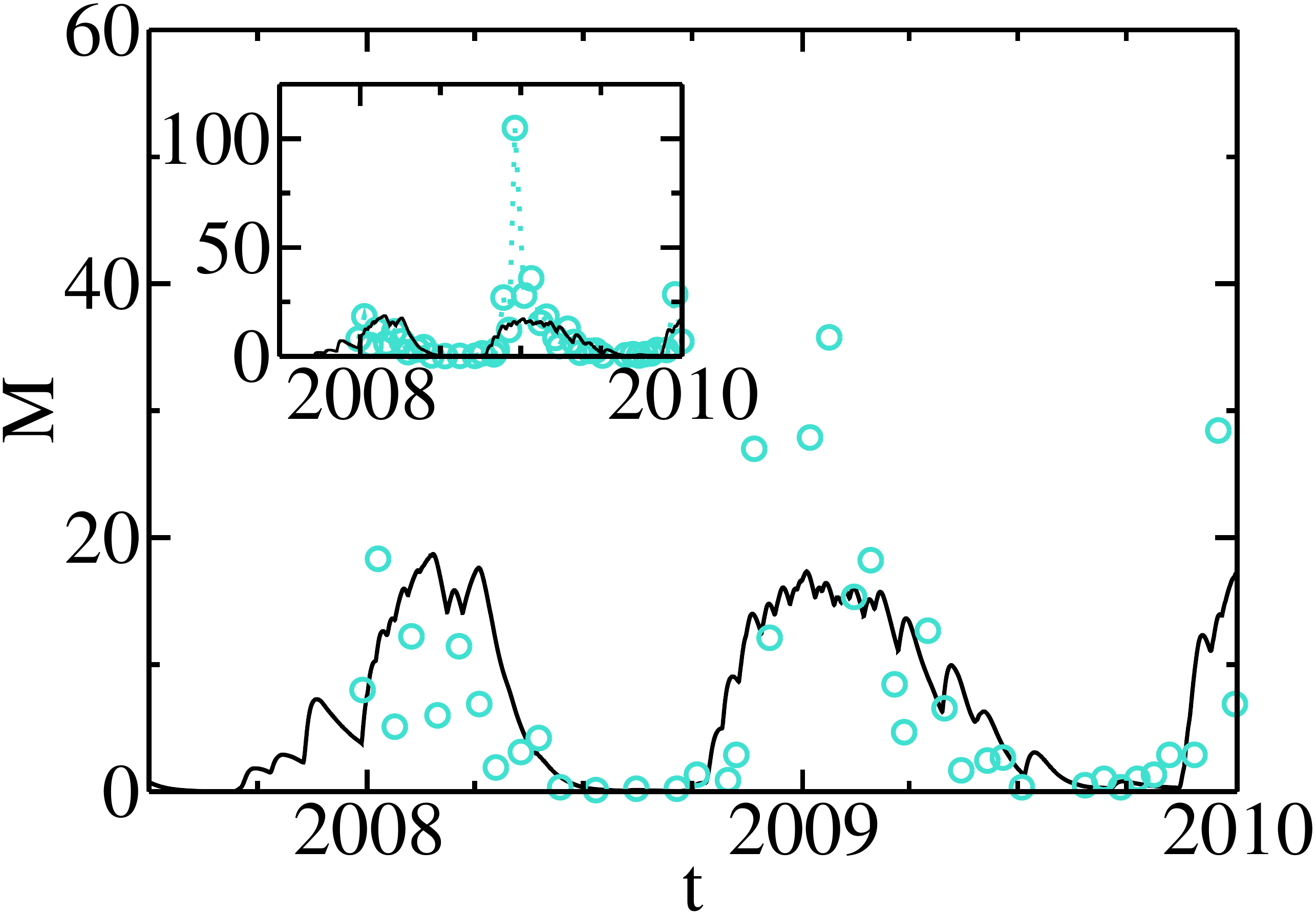}
    \put(80,50){(a)}
  \end{overpic}\hspace{0.5cm}\vspace{0.0cm}
  \begin{overpic}[scale=0.25]{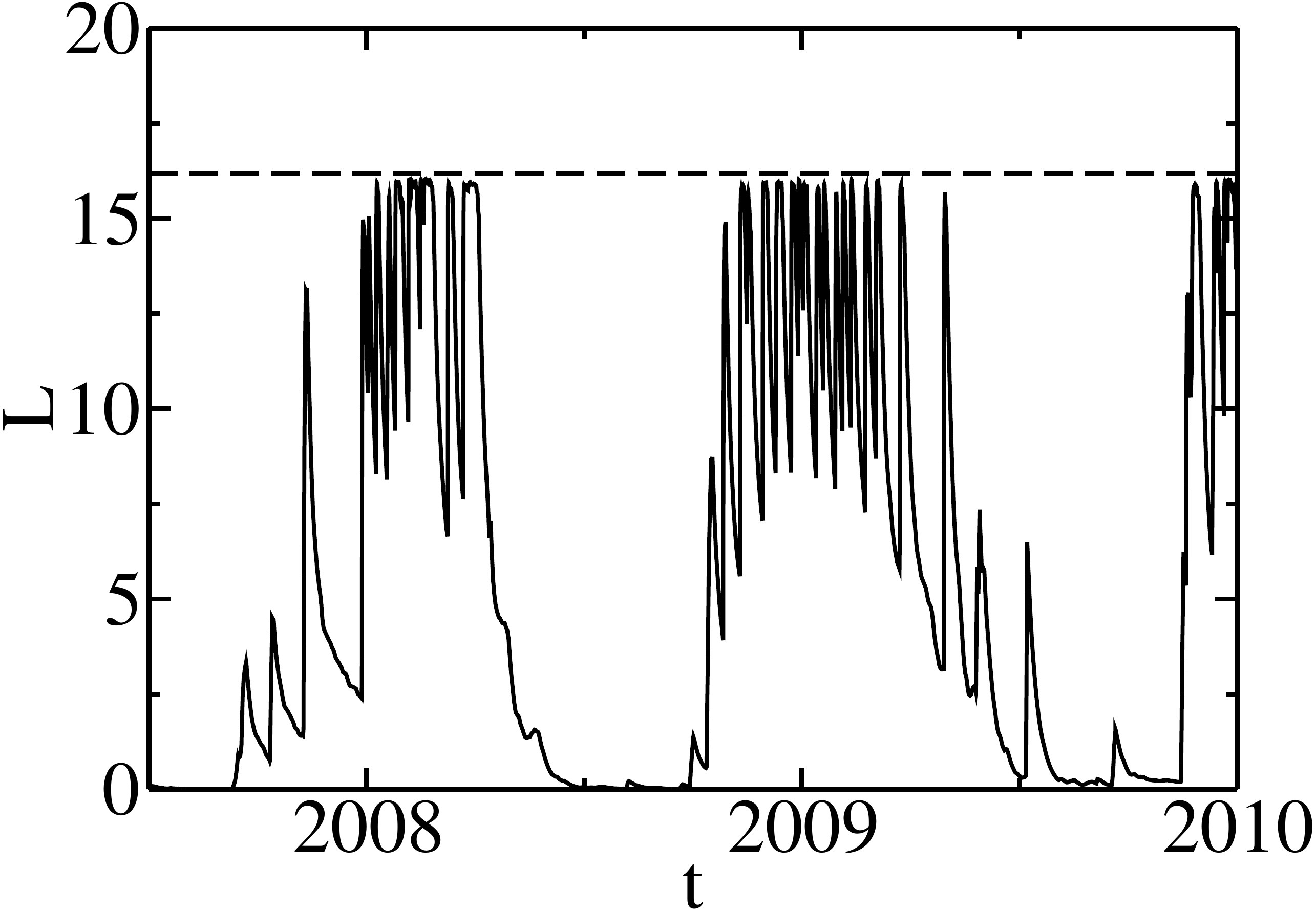}
    \put(15,50){(b)}
  \end{overpic}\hspace{0.5cm}\vspace{0.0cm}
\caption{Panel (a): Evolution of the abundance $M(t)$ of adult
  mosquitoes obtained from the fitted compartmental model. Large tick
  marks correspond to Jan 1. The blue circles represent the number of
  female mosquitoes collected in the city of C\'ordoba, and the black
  line corresponds to the model fit. Inset: main plot on a larger
  vertical scale showing an extremely high abundance data point. Panel
  (b): Evolution of the number of immature mosquitoes obtained from
  Eqs.~(\ref{eq.larv})-(\ref{eq.lamFin}). The dashed line represents
  the carrying capacity, $K_L$. }\label{fig.Ajust}
\end{figure}

Fig.~\ref{fig.Ajust}(a) shows the fit of our model to the data, where
the abundance $M(t)$ is given as the number of female mosquitoes per
night per trap. We observe that the mosquito population, which is
assumed to be proportional to $M(t)$, increases in summer as
expected. Although there are no daily abundance data to compare with,
we remark that our model predicts day-to-day changes in the abundance
of adult mosquitoes $M(t)$ due to fluctuations in temperature,
humidity, and rainfall.

In Fig.~\ref{fig.Ajust}(b) we plot the evolution of the immature
population of mosquitoes $L(t)$ which shows that temperature and
precipitation lead to more abrupt fluctuations in this group than in
the compartment of adult mosquitoes. This was to be expected, since
these climatic variables are directly introduced into the equation of
the population of immature mosquitoes (see Eq.~(\ref{eq.larv})). In
turn, we find that after a rainfall event there is a large increase
in the abundance of this group which frequently approaches the
carrying capacity, leading to a subsequent population decline due to
competition among immature mosquitoes~(\cite{roberts2010larval,suleman1982effects}).

In the following section we study how different rainfall patterns
affect the mosquito population dynamics.

\subsection{Effect of different rainfall regimes on $M_{max}$}
The proposed model of synthetic rainfall allows us to explore the
evolution of mosquito abundance under possible scenarios in which the
weather becomes, for instance, rainier, or with persistent drought
conditions. In this Section we discuss how different rain regimes
influence the mosquito population when it is at its highest (see
Fig.~\ref{fig.Esq1}(c)). To do this, we first assume that the total
rainfall $P(m)$ (see Eq.~(\ref{eq.ArtP})) is equally distributed in
$D(m)$ days (see Eq.~(\ref{eq.ArtD})), the remainder of the days in
the month being rainless.

\begin{figure}[H]
\centering
\vspace{0.5cm}
  \begin{overpic}[scale=0.65]{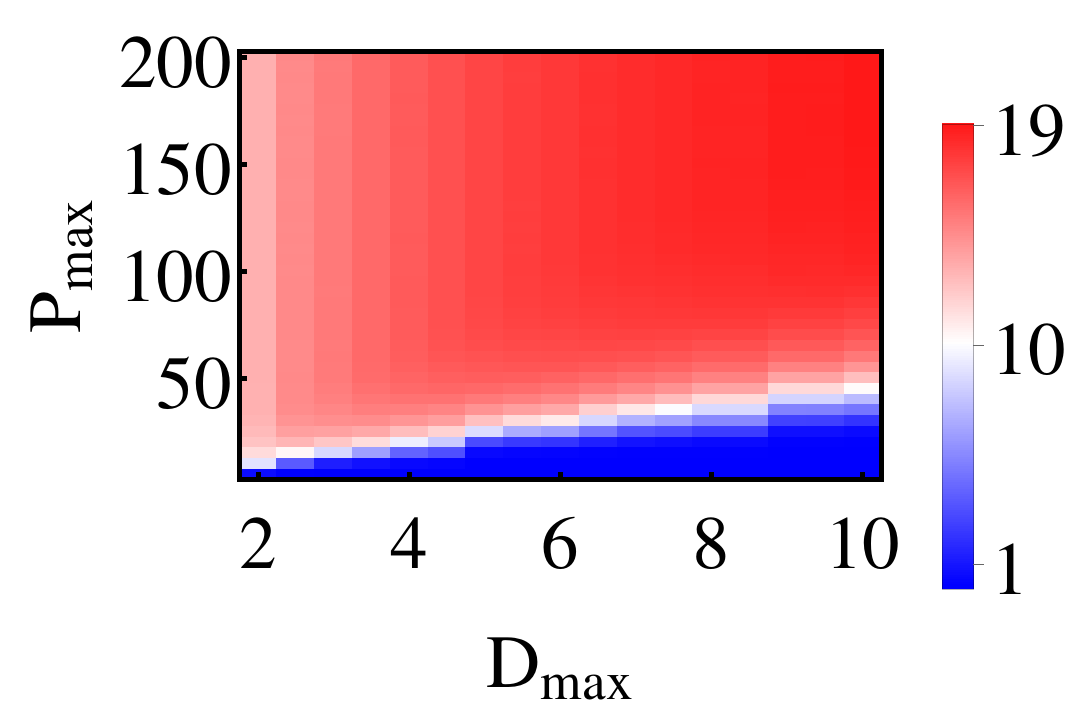}
    \put(70,67){{\bf (a)}}
  \end{overpic}\hspace{0.5cm}\vspace{0.0cm}
    \begin{overpic}[scale=0.25]{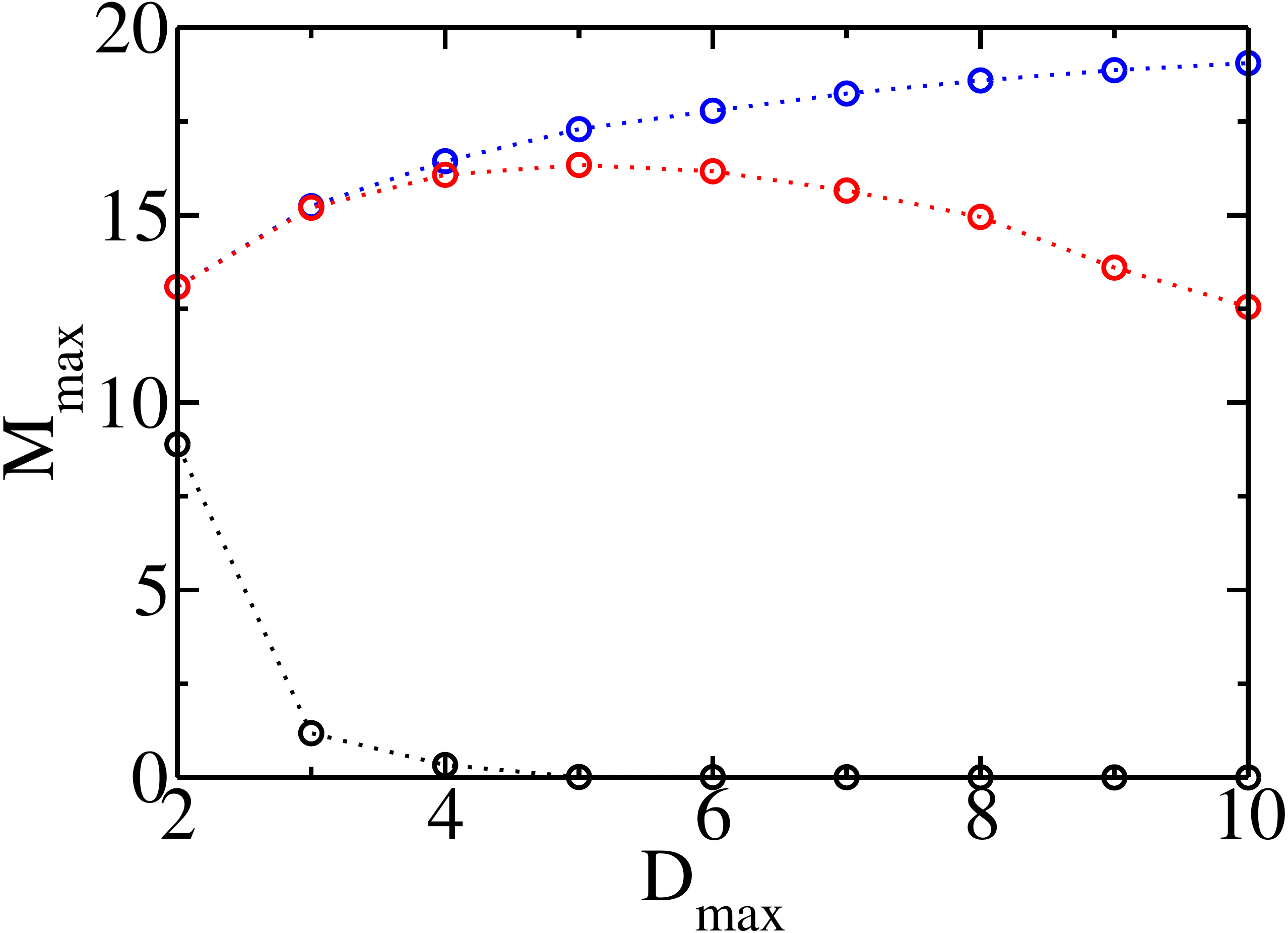}
    \put(79,75){{\bf (b)}}
  \end{overpic}\hspace{0.5cm}\vspace{0.0cm}
\caption{Maximum number of mosquitoes $M_{max}$ in the plane
  $P_{max}$-$D_{max}$ (a) and $M_{max}$ as a function of $D_{max}$ (b)
  for $P_{max}=10$~mm (black), $P_{max}=50$~mm (red) and
  $P_{max}=190$~mm (blue). Dotted lines are to guide the eye. The
  results were obtained from Eqs.~(\ref{eq.larv})-(\ref{eq.lamFin})
  and $10^4$ stochastic realizations.}\label{fig.PhaDiag}
\end{figure}

In Fig.~\ref{fig.PhaDiag}(a) we consider the weather data (temperature and humidity) of the
austral summer season 2008-2009, to show
the influence of $P_{max}$ and $D_{max}$ on the highest peak $M_{max}$
of mosquito abundance, with fixed values of the parameters $P_{min} =
10$~mm and $D_{min} =1$.

From Fig.~\ref{fig.PhaDiag}(a), we note that, for a fixed number of
rainy days $D_{max}$, the highest peak of mosquito abundance increases
as $P_{max}$ grows since, as expected, a greater amount of water
promotes breeding sites for mosquitoes. Similarly, it can be seen from
Figs.~\ref{fig.PhaDiag}(a) and (b) that for $P_{max} \approx 190$~mm,
$M_{max}$ is an increasing function with $D_{max}$, because a higher
frequency of rainfall events provides more opportunities for
mosquitoes to breed. The opposite happens for the lowest level of
precipitation ($P_{max} \approx 10$~mm) as there is very little daily
rainfall in this regime and the accumulated water evaporates
quickly. Interestingly, we find that at moderate precipitation levels
$M_{max}$ has a maximum at an intermediate value of the number of
rainy days. If the rainfall is equally distributed over a few days,
the mosquito population will increase with more rainy days, but,
beyond certain point, the rain becomes too thin to maintain all
breeding sites active and the mosquito population must decrease.
However, as it was mentioned above, the intensity of daily rainfalls
is usually far from uniform~(\cite{Exxon_01}), and recent studies
suggest that the amount of mosquitoes depend on the distribution of
precipitation~(\cite{cheng2016climate,wang2016stage,bomblies2012modeling,bomblies2008hydrology}). Therefore,
in the following we study how heterogeneity in the daily rainfall
affects the dynamics of mosquito abundance.

\begin{figure}[H]
\centering
\vspace{0.5cm}
  \begin{overpic}[scale=0.55]{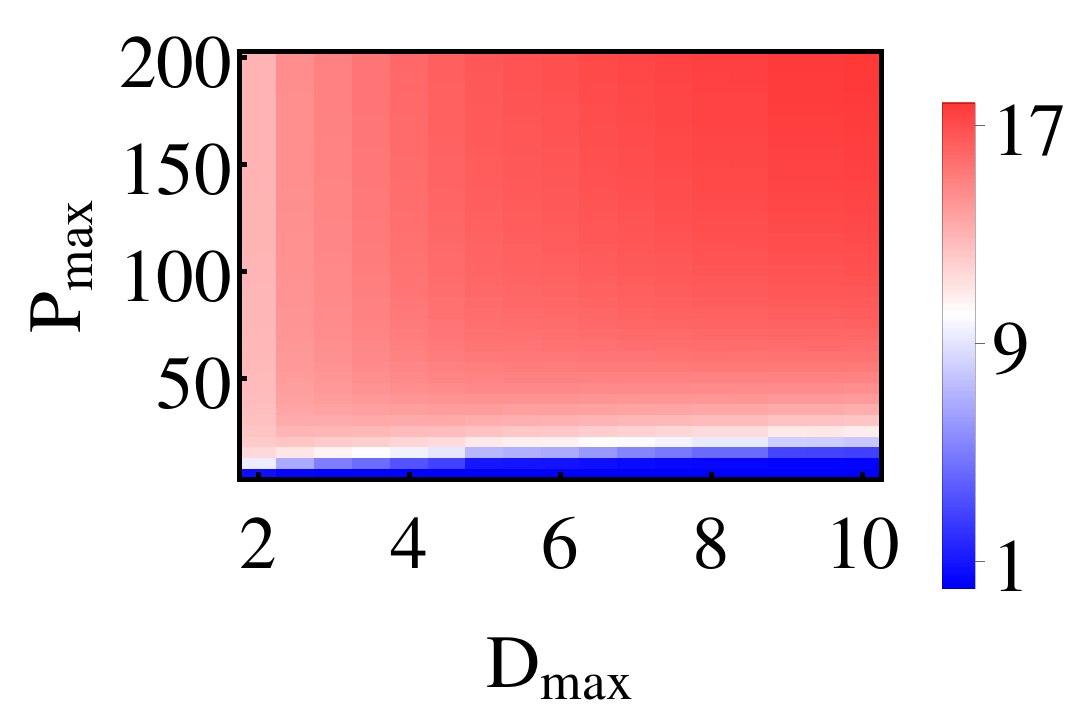}
    \put(90,73){(a)}
  \end{overpic}
    \begin{overpic}[scale=0.23]{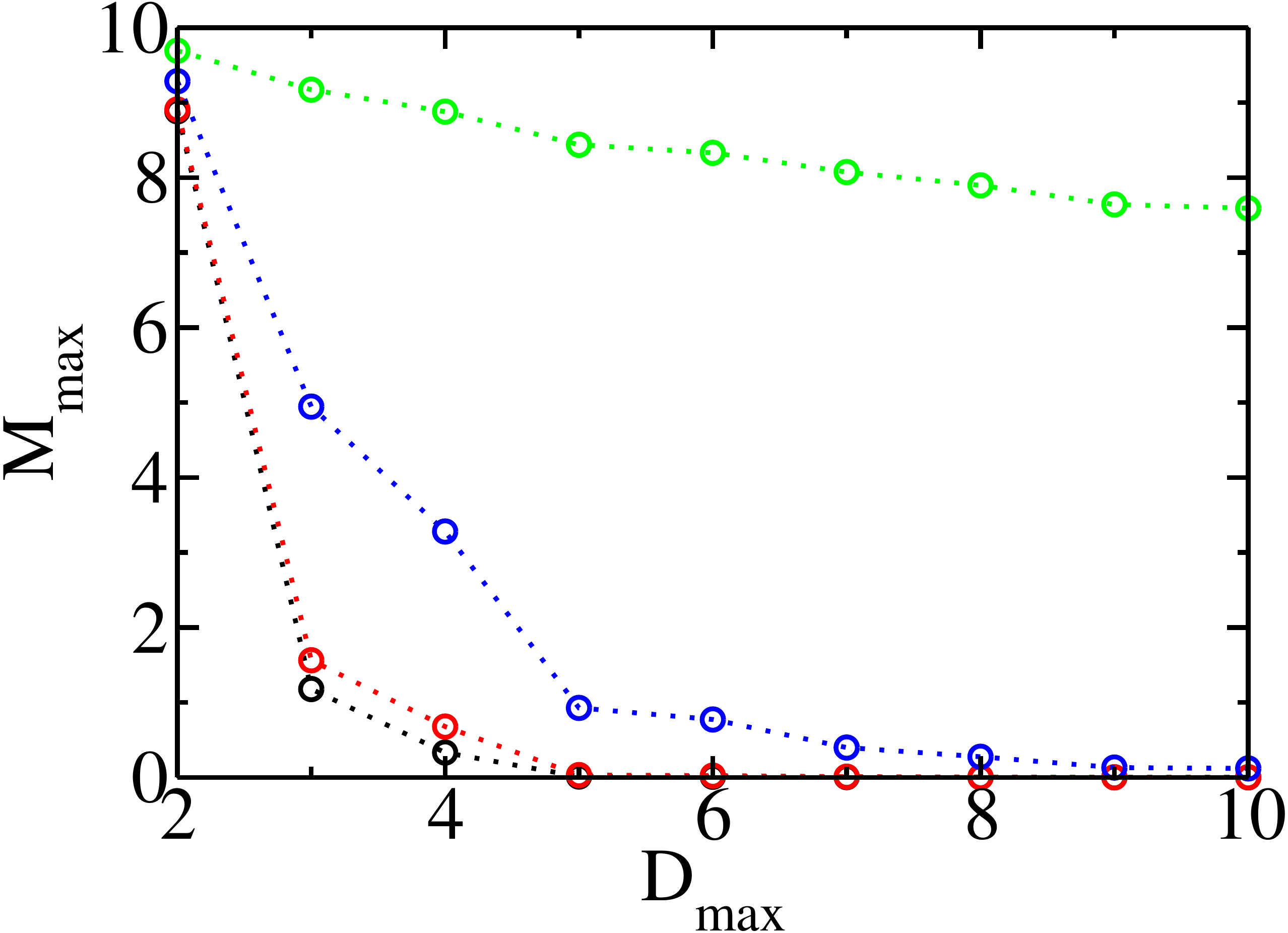}
    \put(90,75){(b)}
    \end{overpic}
 \begin{overpic}[scale=0.23]{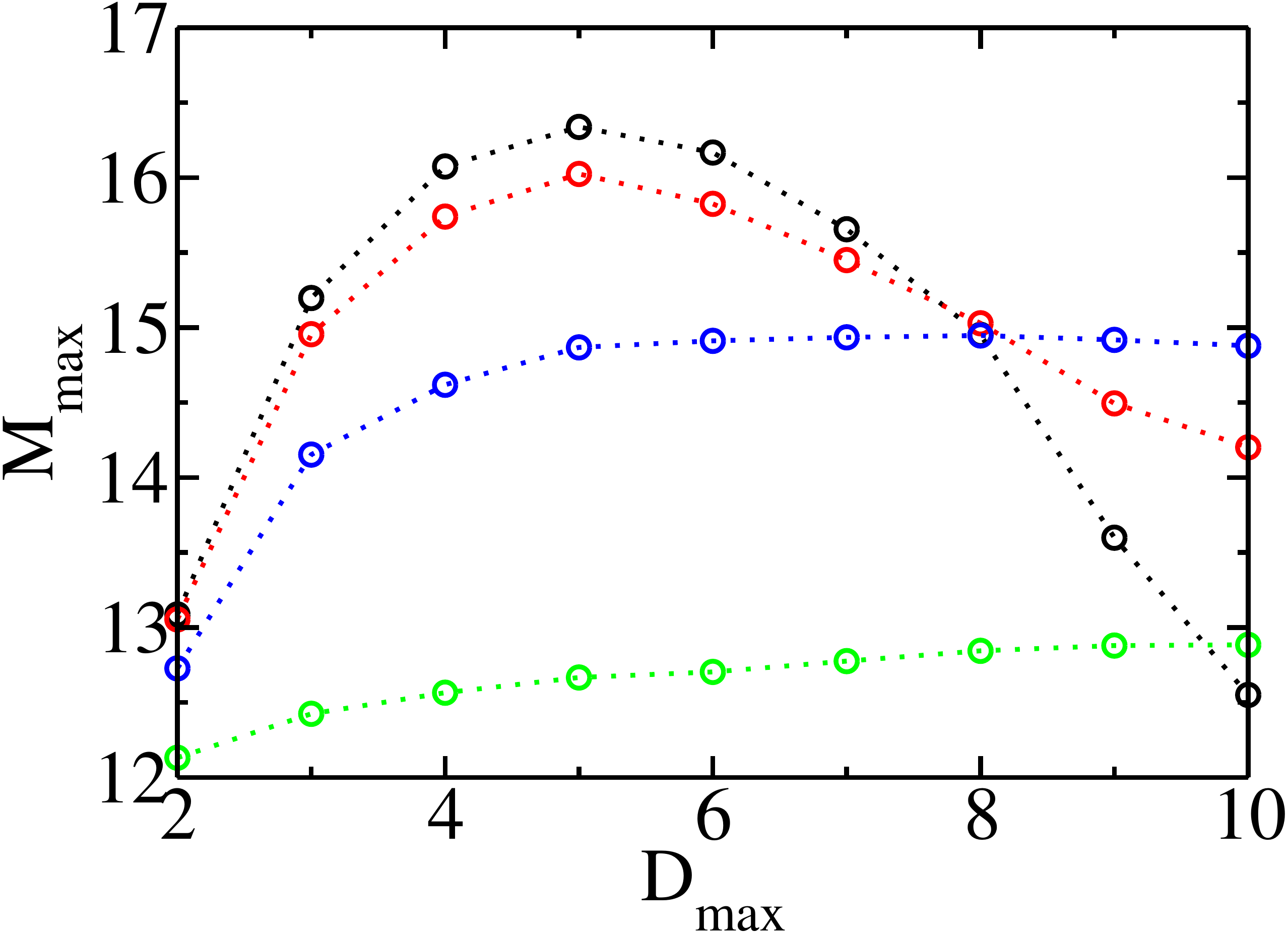}
    \put(20,60){(c)}
  \end{overpic}
 \begin{overpic}[scale=0.23]{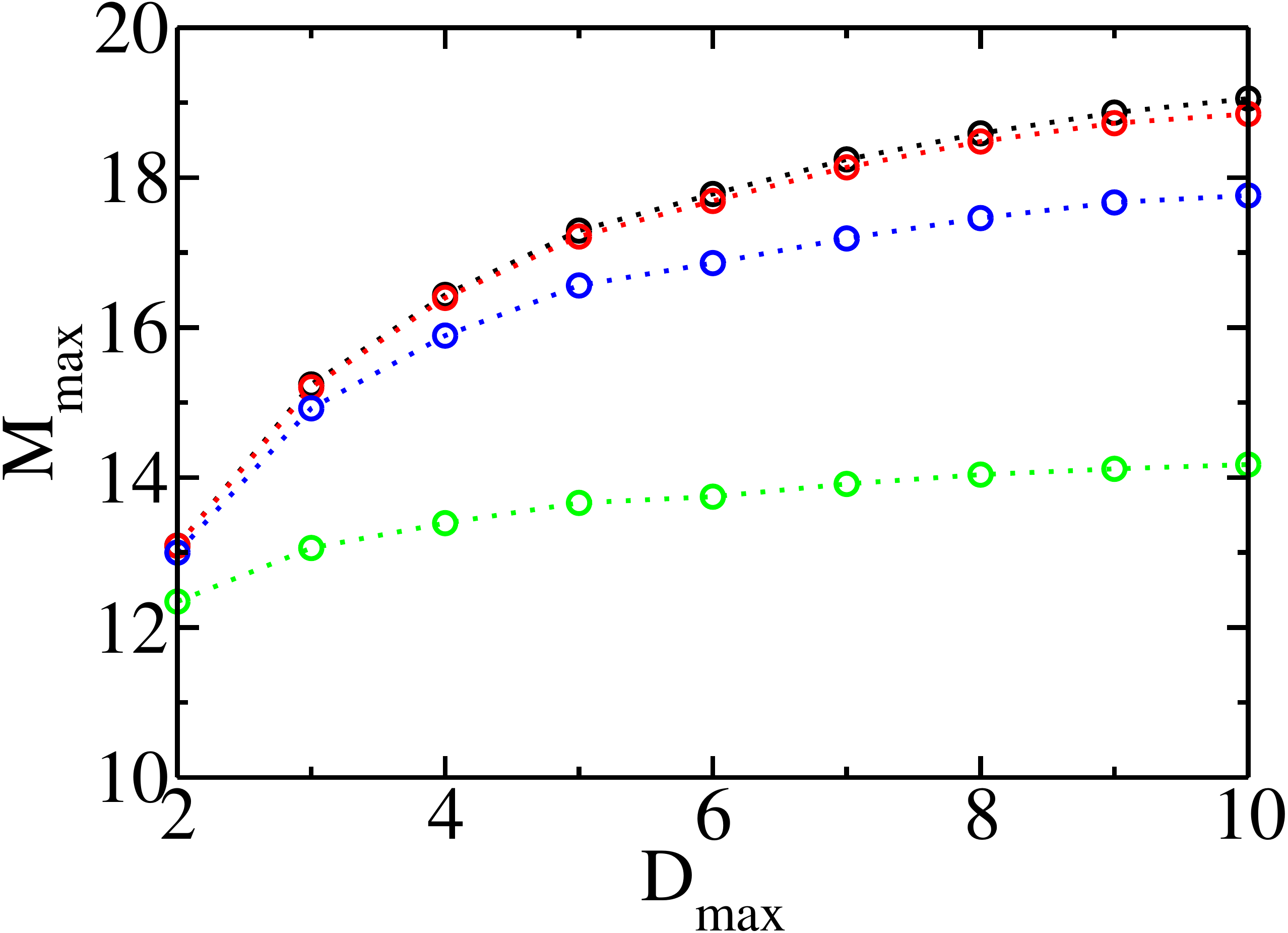}
    \put(20,60){(d)}
  \end{overpic}
\caption{Panel (a): Maximum number of mosquitoes $M_{max}$ in the
  plane $P_{max}$-$D_{max}$ for $\alpha=0.5$ and fixed values of the
  parameters $D_{min}=1$ and $P_{min}=10$~mm. Panel (b): $M_{max}$ as
  a function of the number $D_{max}$ of rainy days in the wettest
  month for $P_{max}=10$~mm for different values of $\alpha$. Rainy
  days have either the same amount of rainfall intensity (black), or
  $\alpha=0.1$ (red), $\alpha=0.5$ (blue), or $\alpha=0.9$
  (green). Panels (c) and (d) show the same as (b) for $P_{max}=50$~mm
  and $P_{max}=190$~mm, respectively. Averaging was made over $10^4$
  realizations of the rainfall time series
  $R(t)$.} \label{fig.AltAlpha}
\end{figure}

In figure~\ref{fig.AltAlpha}, using the weather data of the austral
summer season 2008-2009 (temperature and humidity), we show the peak
mosquito abundance for different values of $P_{max}$, $D_{max}$, and
$\alpha$, as obtained from the
Eqs.~(\ref{eq.larv})-(\ref{eq.lamFin}). For simplicity, we use the
same value of $\alpha$ for each month.

From Figs.~\ref{fig.AltAlpha}(a), (b) and (d), we note that, similarly
to the case of constant rainfall intensity, the abundance $M_{max}$ is
a decreasing (increasing) function with $D_{max}$ for a low (high)
amount of monthly precipitation $P_{max}$. Furthermore,
Figs.\ref{fig.AltAlpha}(b)-(d) show that a higher variability in the
intensity of rainfalls reduces the dependence of $M_{max}$ with
$D_{max}$ with respect to the case of homogeneous rainfall. In
particular, for $P_{max}=10$~mm (see Fig.\ref{fig.AltAlpha}(b)), we
note that $M_{max}$ is higher in the heterogeneous case ($\alpha>0$)
than in the homogeneous one, since a greater variability tends to
confine most of the total monthly precipitation to a few days, in
which there is a higher level of water (see Eq.~(\ref{cpattern}))
available for the immature mosquito population. In contrast, for
$P_{max}=190$~mm (see Fig.\ref{fig.AltAlpha}(d)) the heterogeneity
diminishes the abundance $M_{max}$ with respect to the homogeneous
one. In this case, even if a higher amount of precipitation in a few
days would increase the rate of immature mosquito birth, the effect of the
intense rain days is limited by the threshold $H_{max}$ because the
impact of precipitation saturates for a precipitation higher than
$H_{max}$ ($i.e.$, $\lambda(t)=1$, see Eqs.~(\ref{cpattern})
and~(\ref{eq.lamFin})). Moreover, there are more rainy days with low
precipitation in this regime which further reduces the growth of the
mosquito population. As a consequence, the heterogeneity in the
intensity of rainfall implies that the monthly total precipitation is
a more relevant variable for the prediction of the maximum abundance
of mosquitoes than the number of rainy days.

Another aspect of relevance for the prediction of vector-borne
diseases is the effect of the rainfall in the dry-season (winter) on
the future abundance of mosquitoes in summer. To study this
relationship, we measure the maximum abundance of mosquitoes $M_{max}$
and the timing $\tau_{max}$ at which the peak of mosquito abundance is
reached (see Fig.~\ref{fig.Esq1}(c)), for different values of
$P_{min}$ and $D_{min}$, which correspond to the parameters that
control the intensity and frequency of rainfall in the driest month,
respectively. Here, we keep fixed the parameters $P_{max} =150$~mm and
$D_{max}=10$, and assume that February is always the wettest month of
the year ($m_0=2$, see Eq.~(\ref{eq.ArtP})).
\begin{figure}[H]
\centering
\vspace{0.5cm}
  \begin{overpic}[scale=0.55]{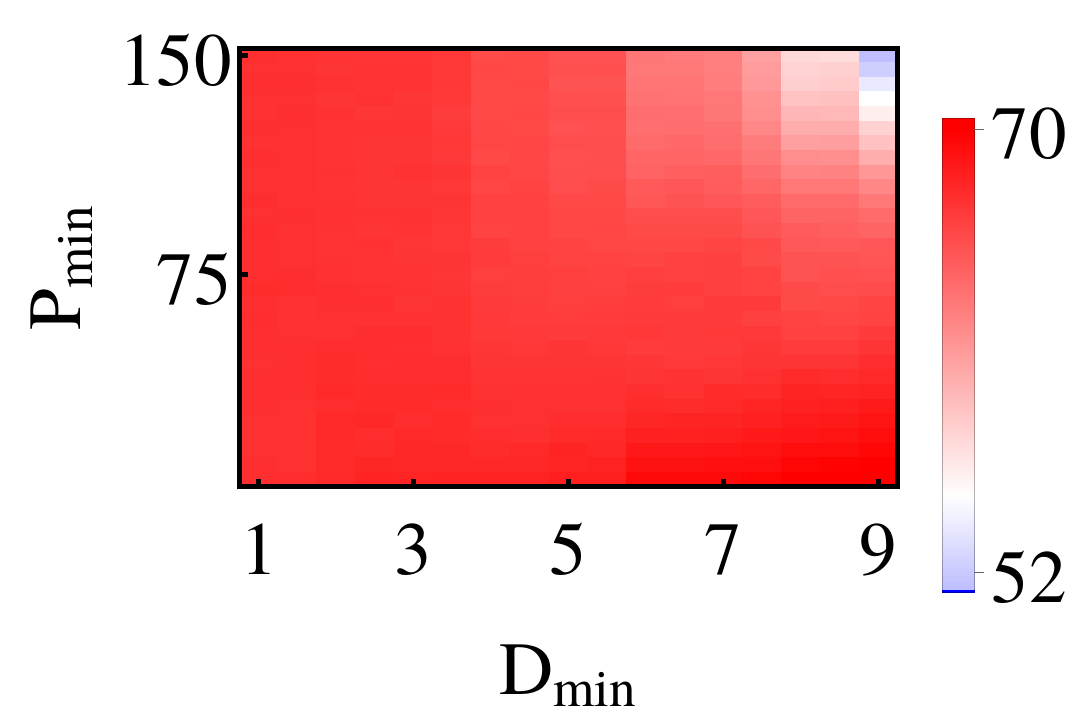}
    \put(20,67){(a)}
  \end{overpic}\hspace{0.5cm}\vspace{0.0cm}
    \begin{overpic}[scale=0.55]{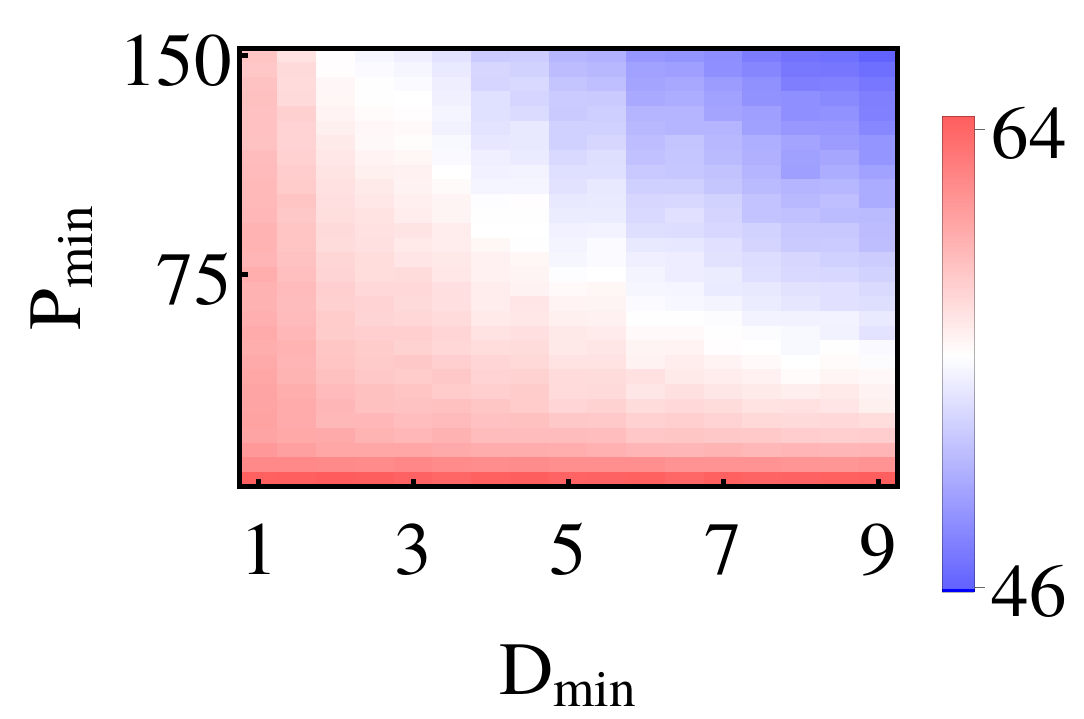}
    \put(20,71){(b)}
    \end{overpic}\hspace{0.5cm}\vspace{0.0cm}
 \begin{overpic}[scale=0.25]{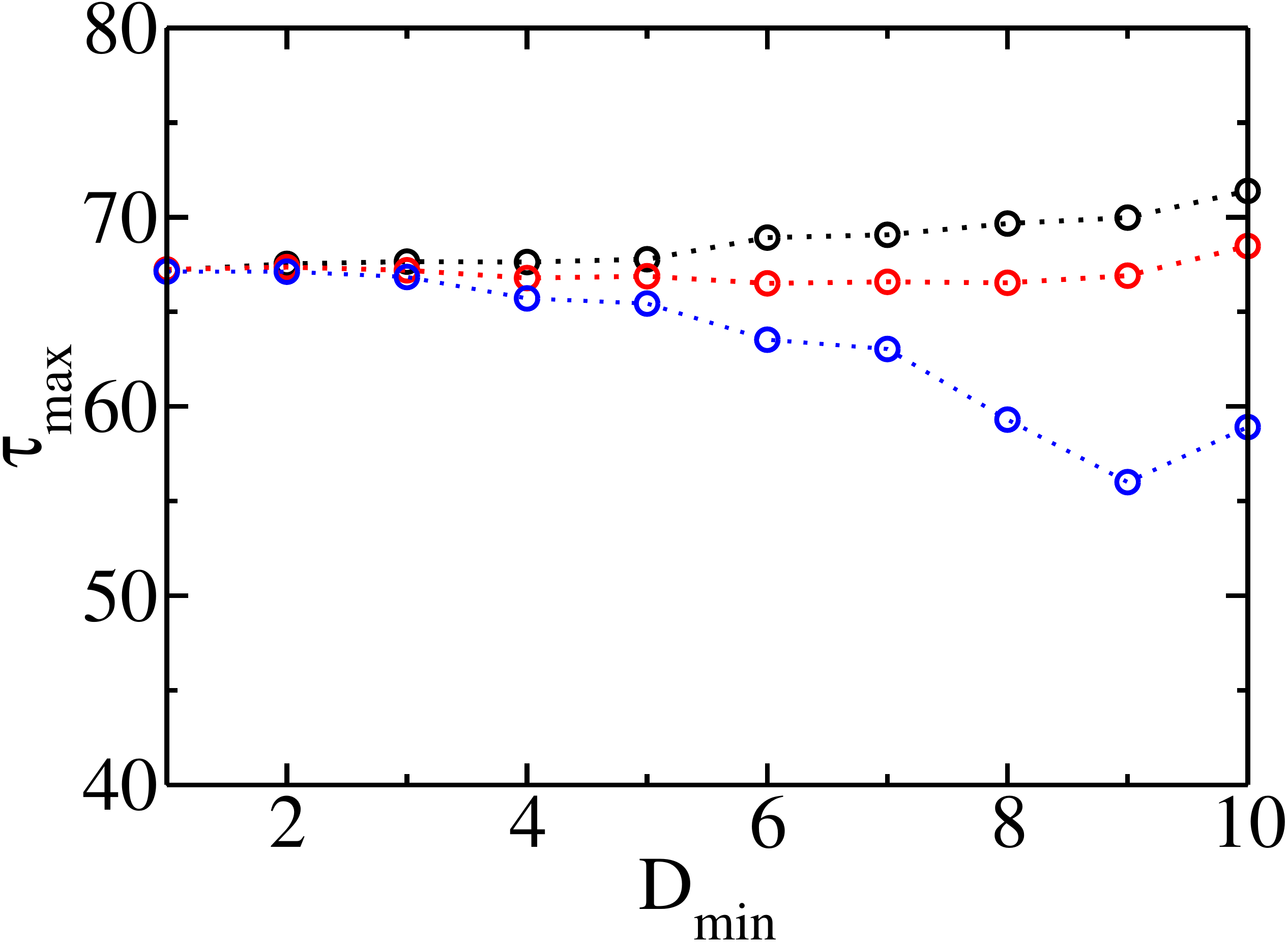}
    \put(20,26){(c)}
  \end{overpic}\hspace{0.5cm}\vspace{0.0cm}
 \begin{overpic}[scale=0.25]{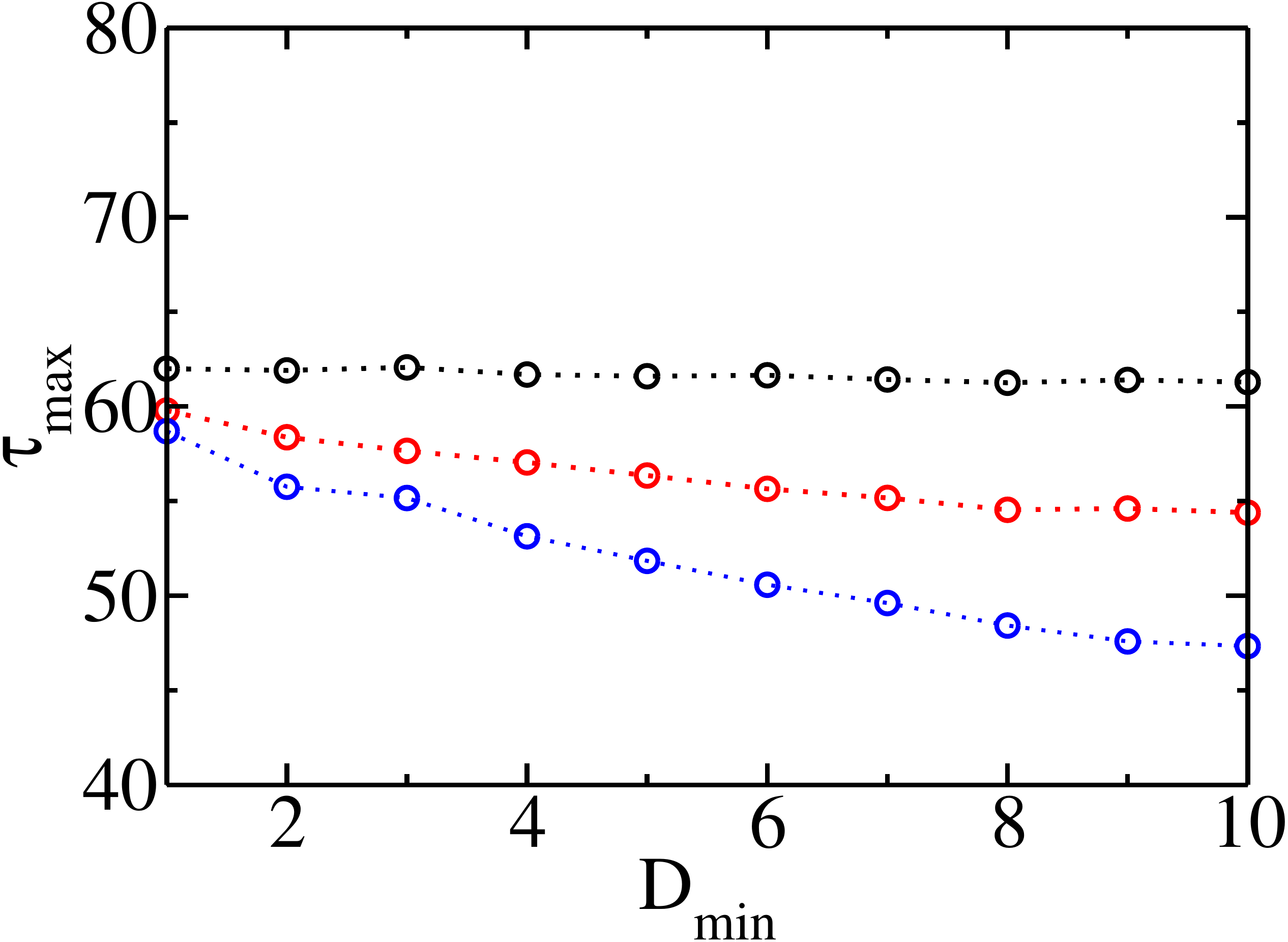}
    \put(20,26){(d)}
 \end{overpic}\hspace{0.5cm}\vspace{0.0cm}
  \begin{overpic}[scale=0.25]{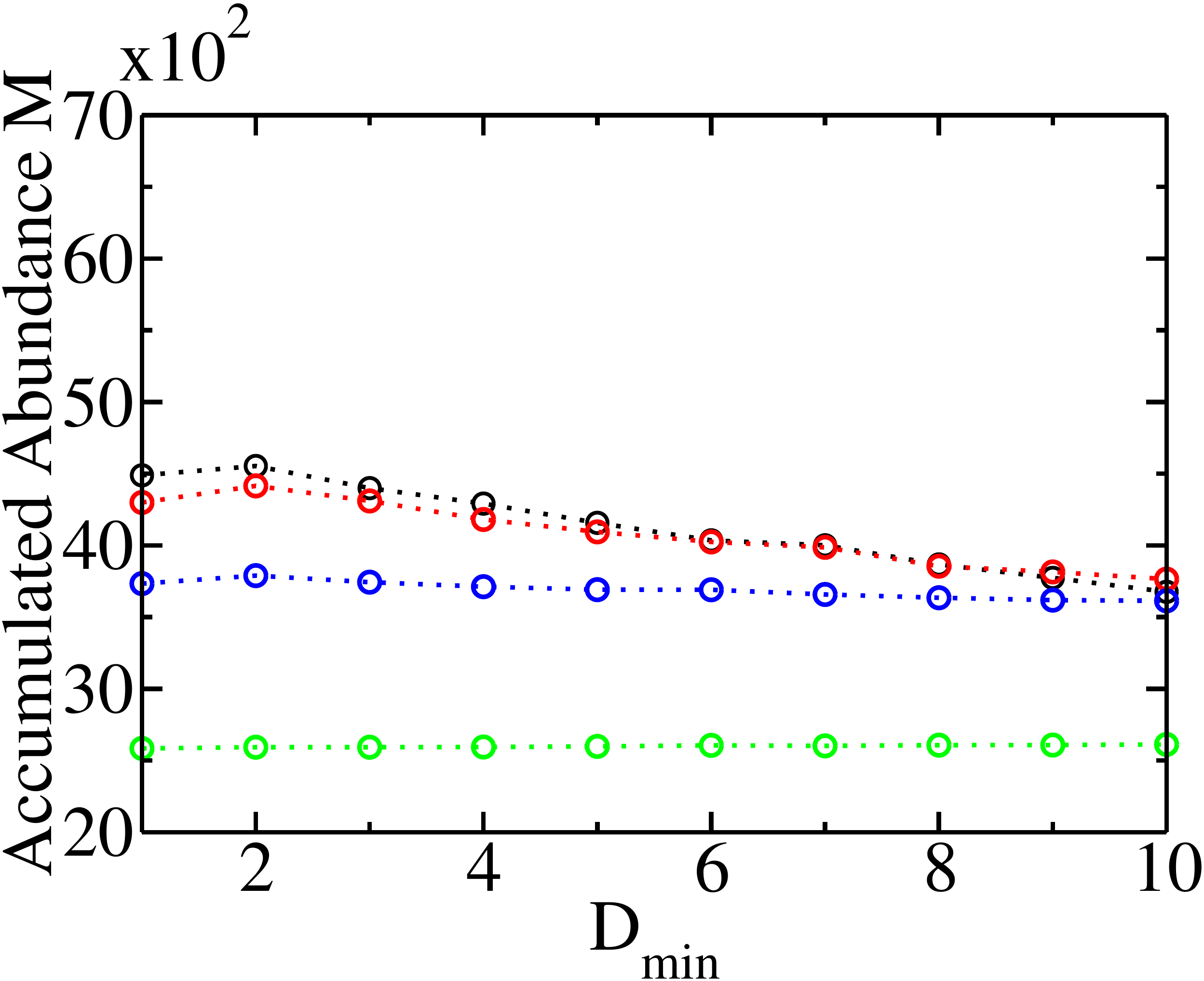}
    \put(20,26){(e)}
  \end{overpic}\hspace{0.5cm}\vspace{0.0cm}
  \begin{overpic}[scale=0.25]{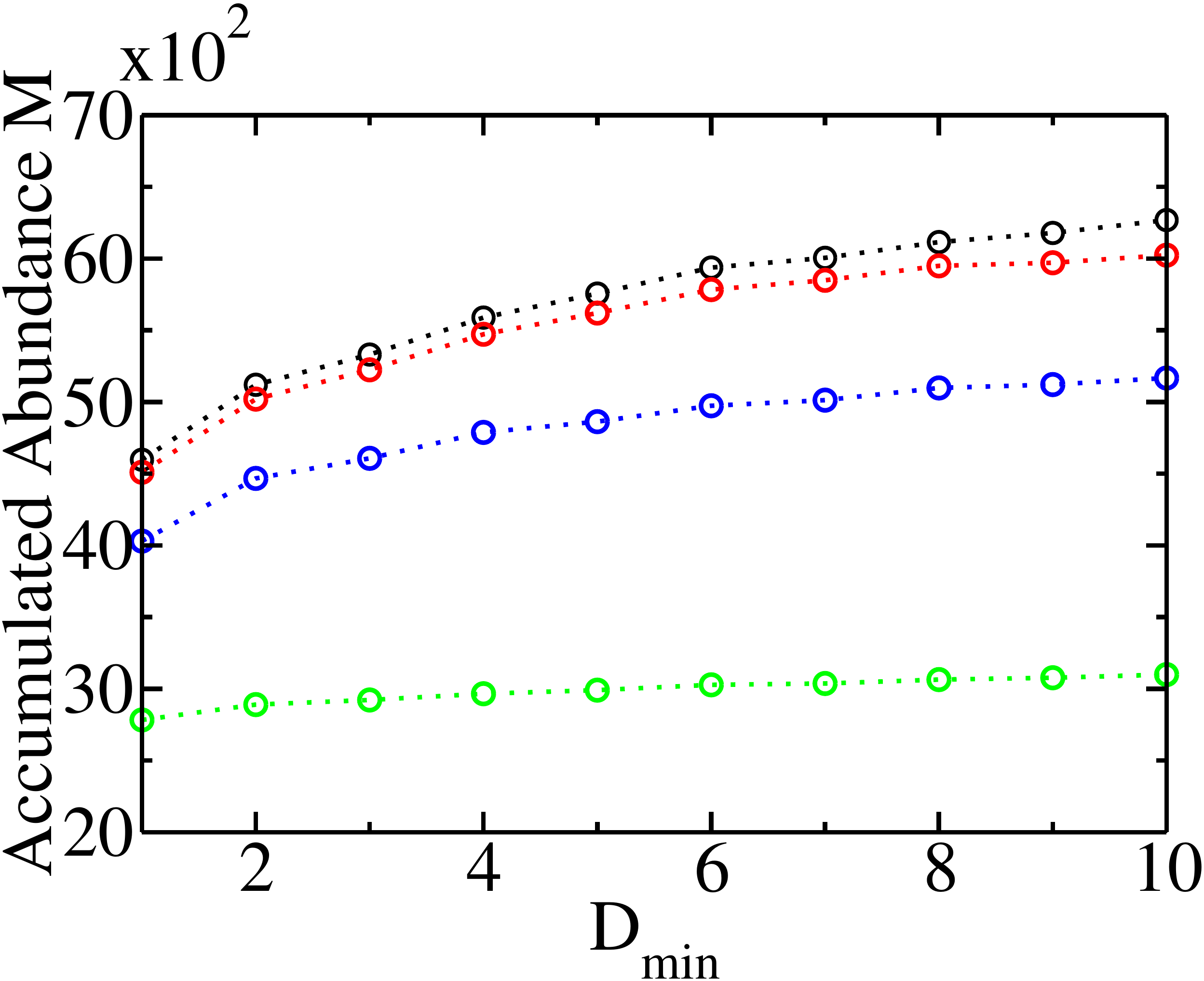}
    \put(20,26){(f)}
  \end{overpic}\hspace{0.5cm}\vspace{0.0cm}
\caption{Panels (a) and (b): Time $\tau_{max}$ (measured in days,
  where $\tau_{max}=0$ corresponds to January 1, 2009) at which the
  peak of mosquito abundance is reached in the plane
  $P_{min}$-$D_{min}$ for a homogeneous intensity distribution of
  rainfalls (a) and $\alpha=0.9$ (b). Panels (c) and (d): $\tau_{max}$
  as a function of the number of rainy days in the driest month,
  $D_{min}$, for a homogeneous intensity distribution of rainfalls (c)
  and $\alpha=0.9$ (d) for various values of $P_{min}$: 10~mm (black),
  50~mm (red) and 130~mm (blue). Panels (e) and (f): accumulated
  mosquito abundance between July 2008 and July 2009 as a function of
  $D_{min}$, with $P_{min}=10$~mm (e) and $P_{min}=130$~mm (f) for:
  homogeneous rainfall intensity (black), $\alpha=0.1$ (red),
  $\alpha=0.5$ (blue) and $\alpha=0.9$ (green). The dotted lines are a
  guide to the eye. The figures were obtained averaging over $10^5$
  realizations of the rainfall time series $R(t)$.}\label{fig.Winter}
\end{figure}
We note from Figs.~\ref{fig.Winter}(a) and (c) that, in the case of a
homogeneous distribution of rainfalls, the time of peak $\tau_{max}$
moves forward for high values of $P_{min}$ and $D_{min}$, because in
this case the rainfalls are regular and abundant throughout the year,
which favors mosquito breeding. However, for the explored values of
$P_{min}$ and $D_{min}$, the position of this peak is in late February
or March, $i.e.$, just after the wettest month of the year. On the
other hand, in a scenario of heterogeneous rainfalls with $\alpha=0.9$
(see Figs.~\ref{fig.Winter}(b) and (d)), $\tau_{max}$ is moved forward
by only approximately 10 days with respect to the homogeneous
intensity case. Correspondingly, for constant values of $\alpha$ (0.1,
0.5 and 0.9) the changes in $P_{min}$ and $D_{min}$ only affect
$M_{max}$ by less than 5\% (not shown here). Therefore these results
suggest that the peak of abundance of mosquitoes and $\tau_{max}$ are
mainly determined by summer weather conditions and the carrying
capacity of the system and not by the intensity and distribution of
precipitation throughout the year. Despite the weak effect of
$P_{min}$ on $M_{max}$, Figs.~\ref{fig.Winter} (e) and (f) show, as
expected, that an increasing value of $P_{min}$ could have a
remarkable effect on the accumulated abundance of mosquitoes (one
measure of which is the time integral of $M(t)$ over the period of
interest), since from $P_{min}=10$~mm to $P_{min}=130$~mm, it could
increase by more than 40\%. However, for the case of a fixed value of
$P_{min}$ and higher values of $\alpha$ we obtain that the accumulated
abundance of mosquitoes diminishes down to a 50\% of the value for a
homogeneous rain distribution. Consequently, the heterogeneity could
help to attenuate the enhancement of the mosquito
population. Therefore, these findings suggest that in order to predict
the total annual abundance, it is not only necessary to take into
account the overall amount of rainfall throughout the year but also
the heterogeneity in daily rainfall intensity.

\section{Discussion}\label{sec.Discussion}
Since projections of climate change~(\cite{nunez2009regional}) suggest
that for the late twenty-first century in regions of South America,
such as C\'ordoba province, the pattern of precipitation will change
towards a regime with rainier autumns and an increase in extreme
events, it is crucial to study how this variation would affect the
mosquito abundance.

In this paper we studied the effects of the total intensity,
number of rainy days and heterogeneity of rainfall on the mosquito
population. We found that for a regime with a low total rainfall, the
abundance of mosquitoes is a decreasing function with the number of
rainy days, while for a high total rainfall regime it is an
increasing function of this number. Interestingly, for an intermediate
precipitation regime, we found that there is a halfway number
$D_{max}$ of rainy days for which $M_{max}$ is optimized. Since
$P_{max}$ is fixed, fewer rainy days would imply dry intervals,
leading to a lessening of the mosquito abundance.  If the number of
rainy days exceeds the optimal value of $D_{max}$, a considerable
fraction of the rainwater resulting from the typically meager rainfall
would disappear due to evapotranspiration, again leading to a
reduction of the mosquito abundance.

In order to study the effect of the heterogeneity in the daily
rainfall, we used a fracturing process that keeps constant the total
amount of monthly precipitation. We observed that a higher
heterogeneity reduces the dependence of $M_{max}$ on the number of
rainy days. However, an increasing variability favors the mosquito
production in the low rainfall regime, while the opposite behavior
takes place in the case of high precipitation $P_{max}$. Therefore, if
climatic models predict the intensification of storms, but not an
increase in the total amount of precipitation, our model predicts that
the enhancement of mosquito abundance would be more significant in
semiarid areas than in humid climates.

Finally we study the effect of an increasing amount of rainfall in the
dry season on the mosquito abundance dynamics, obtaining that high
precipitation throughout the year does not significantly alter the
maximum abundance or the time at which this peak occurs, but it could
notably increase the accumulated abundance of mosquitoes. However, we
also observed that a regime with a higher variability of rainfall
intensity could reduce this increase.

While our model captures multiple relationships between rainfall and
mosquito population, additional extensions could be considered. For
instance, there is evidence that rainfalls reduce the immature
population in the short term due to flushing of breeding
sites~(\cite{gardner2012weather,strickman1988rate}) and affect the
bacterial concentration used as food by mosquito
larvae~(\cite{chaves2011weather}); therefore, it would be
interesting to study the relevance of these effects on the dynamic of
mosquito population.

We think that our findings could be used as support and reference
guidance for the assessment of the influence of different rainfall
regimes on the mosquito population dynamics, using the weather data
for any specific region. Such an assessment would impact positively on
our ability to make predictions for the spread of various possible
arboviruses. It is also known that rainfall could have a substantial effect
on insecticide residence times~(\cite{allan2009environmental}). The
model presented here can be used to optimize the efficacy of mosquito
control campaigns, using temperature and rainfall data to select the
best times for the application of population reduction procedures.

\appendix
\counterwithin{figure}{section}
\counterwithin{table}{section}
\renewcommand{\thefigure}{\Alph{section}.\arabic{figure}}
\renewcommand{\thetable}{\Alph{section}.\arabic{table}}
\section{Calibration}\label{AppCalib}
The Metropolis-Hastings (MH) algorithm is a stochastic optimization
tool for fitting statistical models to data that has been used in
cosmology~(\cite{christensen2001bayesian,christensen2003metropolis,lewis2002cosmological}),
epidemiology~(\cite{merler2015spatiotemporal}), and in the study of
mosquito population dynamics~(\cite{marini2016role}). This algorithm
allows us to estimate the unknown values of some parameters $\Theta$
($\Theta$ represents either a single parameter or a parameter set), by
means of a stochastic search in the parameter space that generates a
sequence or chain $\Theta^{(i)}$, where $i$ represents the step number
of the MH algorithm~(\cite{bonamente2013statistics}). Each value of this
chain is sampled from a proposal distribution and accepted with a
probability $\sigma$ defined by an acceptance function, which depends
on the likelihood function of the observations.

In our model we estimate the parameters $\beta_L$, $H_{max}$,
$H_{min}$ and $K_L$ using a MH algorithm and the data for adult
mosquito abundance from C\'ordoba city in the period
2008-2009~(\cite{batallanthesis,batallan2015st}). We propose that the likelihood of the
observations is given by
\begin{eqnarray}
L=\prod_{j=1}^{n}p(x_j(H_{max},H_{min},K_L,\beta_L);k_j),
\end{eqnarray}
where $n$ is the number of data points and
$p(x_j(H_{max},H_{min},K_L,\beta_L);k_j)$ is the probability to
observe the abundance $k_j$ of mosquitoes obtained from the data. Here
we assume that $p(\cdot)$ follows a Poisson distribution whose mean
$x(H_{max},H_{min},K_L,\beta_L)$ is the number of adult mosquitoes
predicted by Eqs.~(\ref{eq.larv})-(\ref{eq.lamFin}).

The MH algorithm implemented in this paper has the following
steps:
\begin{itemize}
\item Step 1: Initialize the starting value of the parameters
  $\Theta^{(i=0)}$, using a uniform distribution in order to avoid
  favoring any initial value.
\item Step 2: Generate a new sample of the parameters, $\Theta^{New}$
  starting from a proposal distribution that indicates a candidate for
  the next sample value. To ensure that the new values of the
  parameters are positive, we use as a proposal distribution a
  log-normal density which has a mean equal to the logarithm of the
  current value parameter and constant variance $\delta$. The value of
  this variance is chosen in order to guarantee an acceptance rate
  between 10\% and 30\% in the burn-in period.
\item Step 3: accept the new candidate $\Theta^{New}$ with probability $\sigma$:
  \begin{eqnarray}
   \sigma &=& min\bigg\{1,\frac{L(\Theta^{New})}{L(\Theta^{(i)})}\bigg\}.
  \end{eqnarray}
\item Step 4: repeat steps 2 and 3 until convergence is reached.
\end{itemize}
We perform $2.10^6$ iterations and check convergence by visual
inspection of the chain $\Theta^{(i)}$. In order to construct the
posterior distribution of the parameters, we discard the first $10^5$
iterations as a burn-in and we only keep every $20^{th}$ sampled value
of the remaining iterations to reduce autocorrelation within successive
samples. Finally, the values of the parameters that we will use in our
model are the averages of the medians of the posterior distributions
obtained from 5 different initial conditions (see step one of the MH
algorithm).

\begin{figure}[H]
\centering
\vspace{0.0cm}
  \begin{overpic}[scale=0.25]{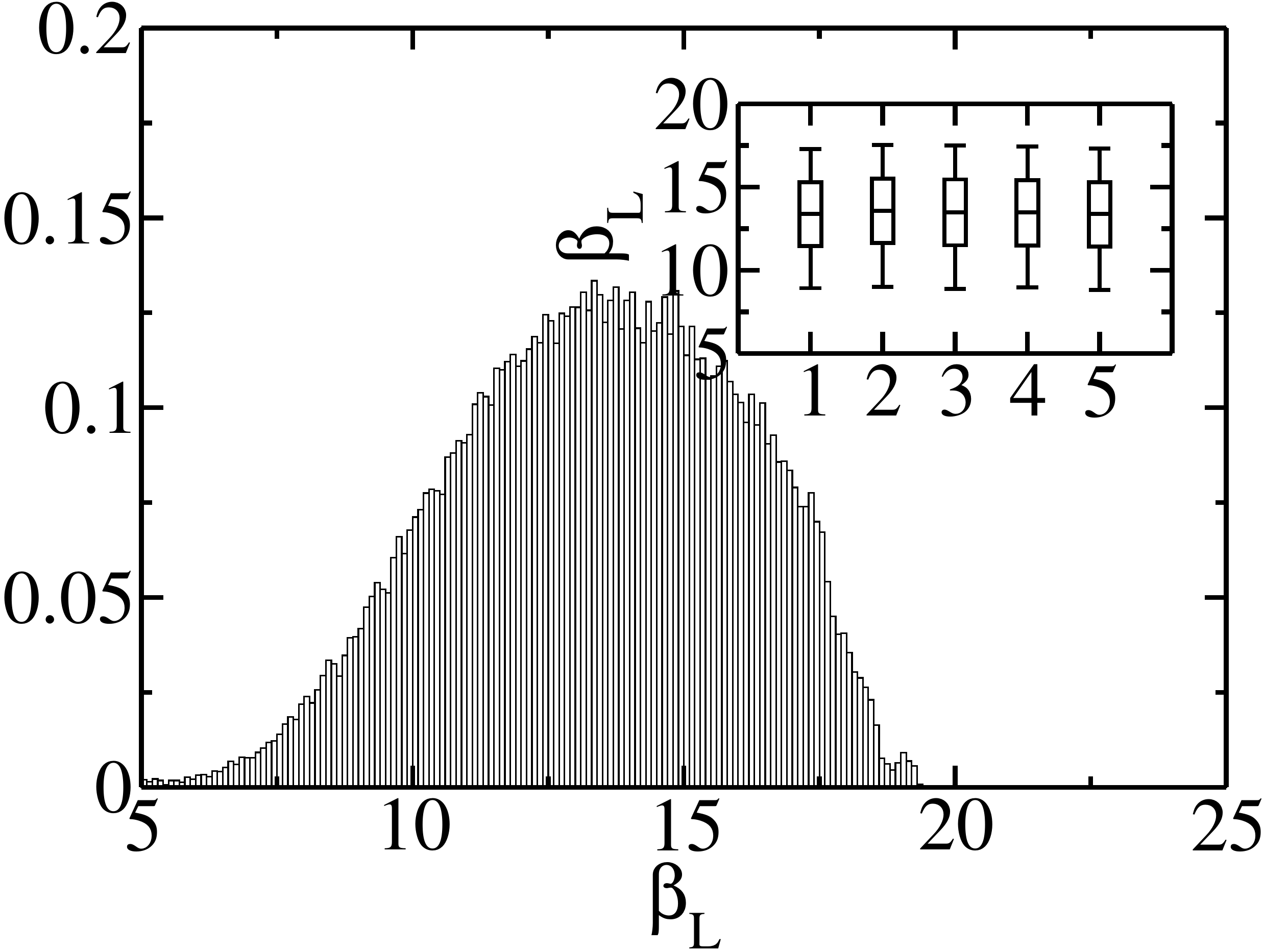}
    \put(15,50){{\bf{(a)}}}
  \end{overpic}\hspace{0.0cm}\vspace{0.0cm}
  \begin{overpic}[scale=0.25]{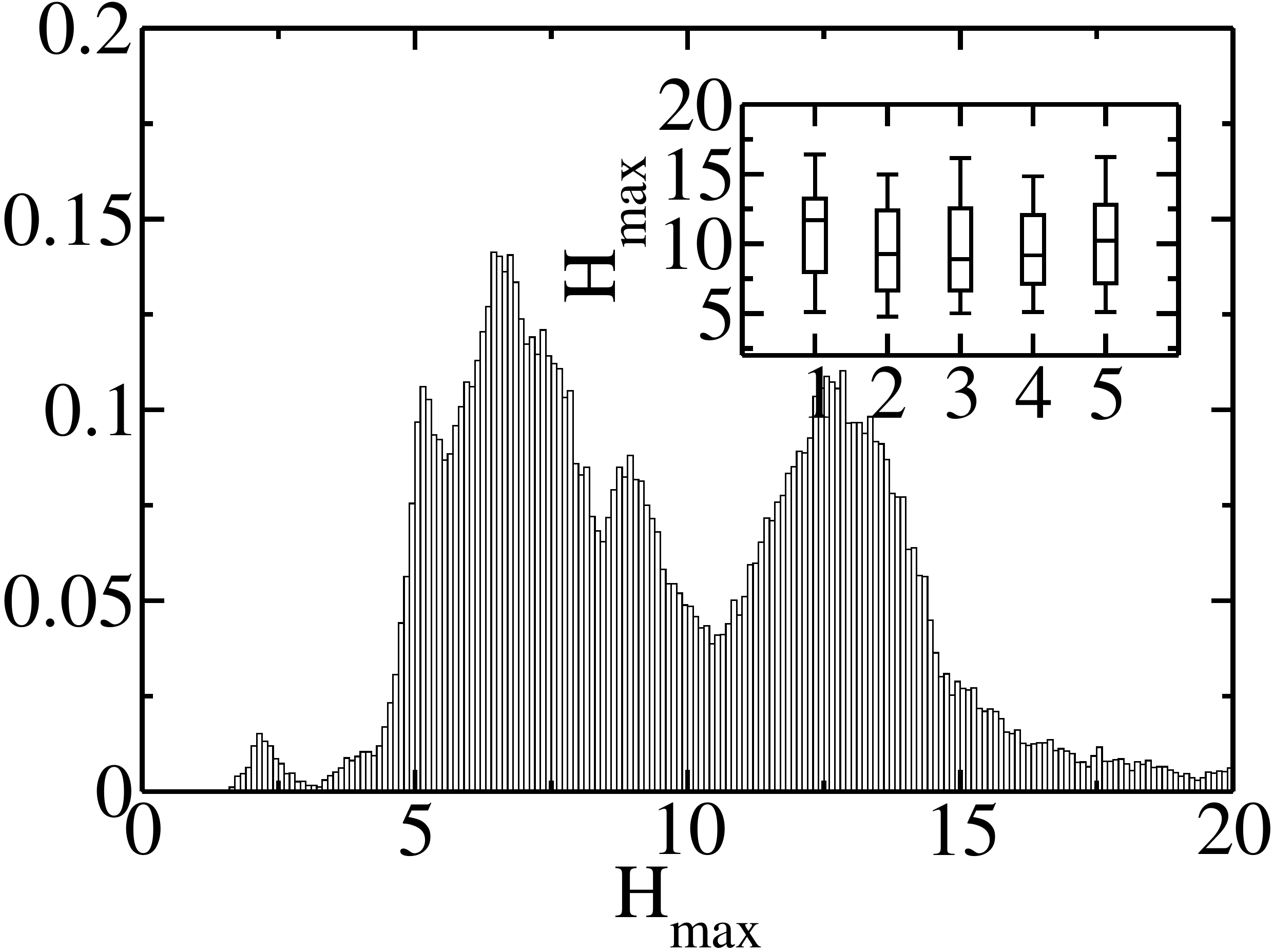}
    \put(15,50){{\bf{(b)}}}
  \end{overpic}\hspace{0.0cm}\vspace{0.0cm}
  \begin{overpic}[scale=0.25]{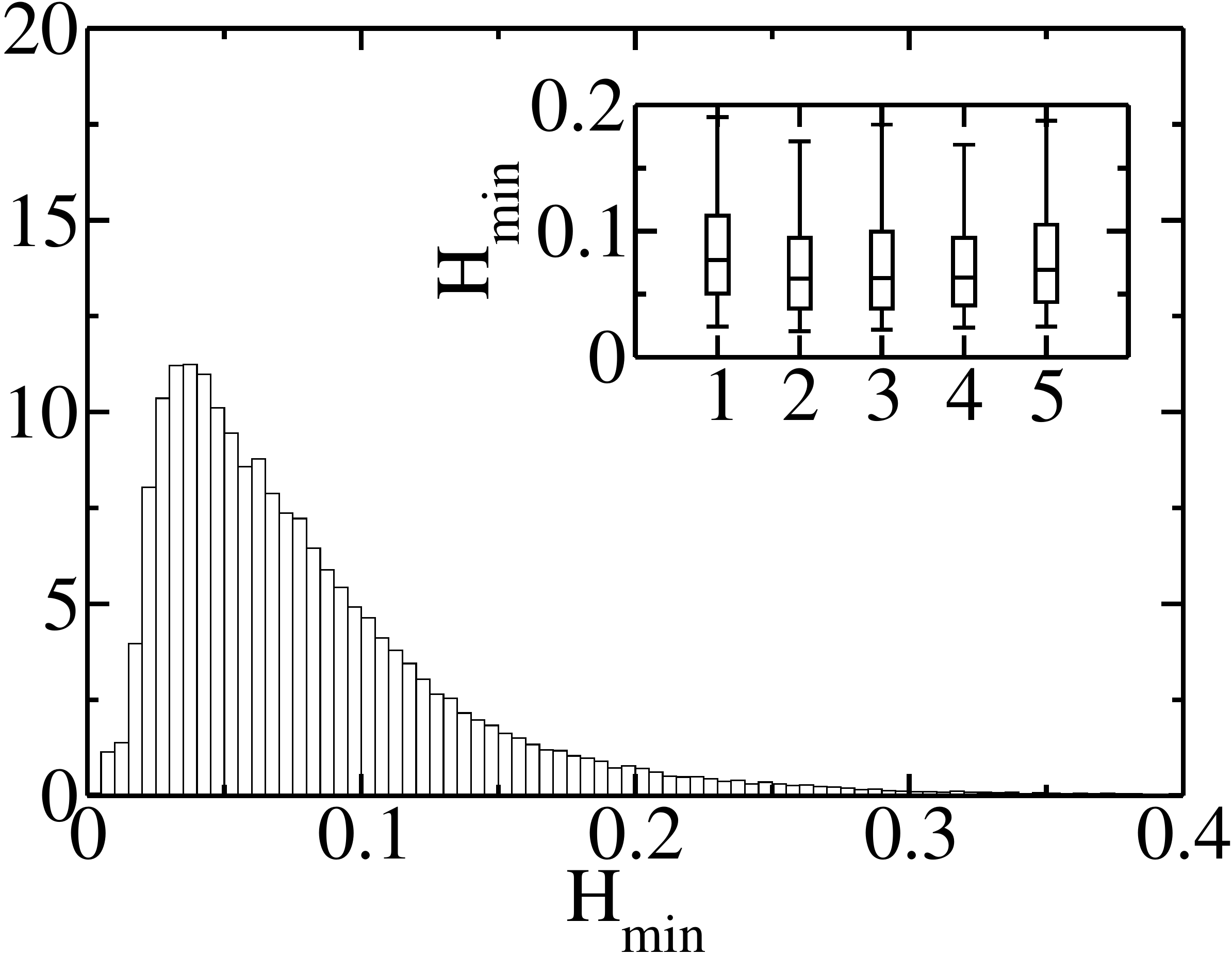}
    \put(15,50){{\bf{(c)}}}
  \end{overpic}\hspace{0.0cm}\vspace{0.0cm}
  \begin{overpic}[scale=0.25]{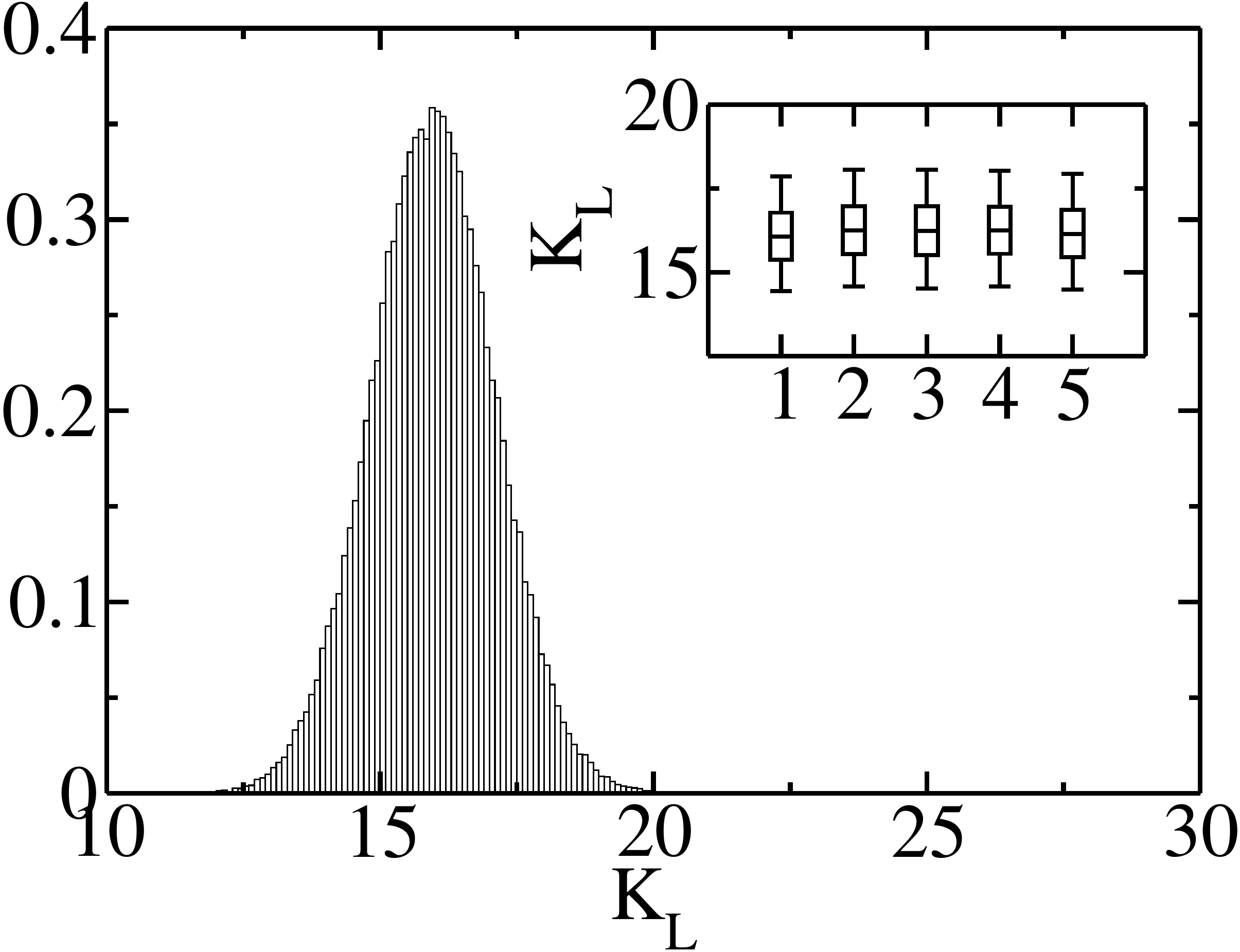}
    \put(15,50){{\bf{(d)}}}
  \end{overpic}\hspace{0.0cm}\vspace{0.0cm}
\caption{Posterior distribution obtained in the steady state of the MH
  algorithm from $2.10^6$ iterations, using $10^5$ iterations as a
  burn in for the parameters $\beta_L$ (a), $H_{max}$ (b), $H_{min}$
  (c), and $K_L$ (d). In the insets we show the boxplot obtained from
  the posterior distribution in the steady state for 5 different
  initial conditions (see step one of the MH
  algorithm).}\label{fig.AppMH}
\end{figure}

Fig.~\ref{fig.AppMH} shows the posterior distribution obtained for the
parameters $\beta_L$, $H_{max}$, $H_{min}$ and $K_{L}$. We note that
all of these distributions are unimodal, except for $H_{max}$.
Although in this paper we set $H_{max}=9.86$~mm, since it is the
average value of the median obtained from the MH algorithm, we also
check our model for $H_{max}\approx 7$~mm and $H_{max}\approx 13$~mm
which are the positions of the highest peaks of the posterior
distribution (see Fig.~\ref{fig.AppMH}(b)). For these cases, our
results presented in Section~\ref{secResul} do no qualitatively
change.

\section{Fracturing process}\label{AppFT}
Fracturing process (FT) is a stochastic iterative process which
generates a finite sequence of numbers with the property that their
sum is always a
constant~(\cite{borgos2000partitioning,finley2014exploring}). From a
geometrical point of view, this method consists of partitioning an
interval of length $\ell$ in a number $D$ of subintervals or segments,
with the property that the sum of their lengths is always
$\ell$. Following ~\cite{finley2014exploring}, the FT process
starts with an interval or segment of length $\ell$ which is split
into two subintervals of lengths $\widetilde{\ell}$ and
$\ell-\widetilde{\ell}$, where $\widetilde{\ell}$ is a stochastic
variable generated by the following function:
\medskip
%\[\nonumber
\begin{eqnarray}\label{alphaFTT}
\widetilde{\ell}=\ell \times \left\{%
\begin{array}{ll}
\rho \frac{1-\alpha}{\alpha} &\;\;\;\; \text{if}\;\;\;\; 0\leqslant \rho <\frac{\alpha}{2} \\
\frac{1}{2}+\left(\rho-\frac{1}{2}\right)\frac{\alpha}{1-\alpha} &\;\;\;\; \text{if}\;\;\;\; \frac{\alpha}{2}\leqslant \rho\leqslant 1-\frac{\alpha}{2} \\
1-(1-\rho)\frac{1-\alpha}{\alpha} &\;\;\;\; \text{if}\;\;\;\; 1-\frac{\alpha}{2} < \rho \leqslant 1 \\
\end{array}%
\right.
\end{eqnarray}

%\]
\medskip
Here $\rho$ is a uniform random variable and $\alpha \in (0,1)$ is a
parameter that controls the average length $\widetilde{\ell}$. In a
second step, the partition function, Eq.~(\ref{alphaFTT}), is applied
again on the intervals resulting from the previous step, generating a
total of 4 intervals. This procedure is repeated until the required
number of intervals is reached \footnote{For example, in order to
  generate a total of five subintervals, three iterations of the
  fracturing process must be performed: step 1) the initial interval
  is divided into two subintervals, step 2) each of the previous
  subintervals is divided into two parts, and finally step 3) one
  randomly chosen subinterval of the previous step is split into two
  subintervals.}.

\begin{figure}[H]
\centering
\vspace{0.5cm}
  \begin{overpic}[scale=0.65]{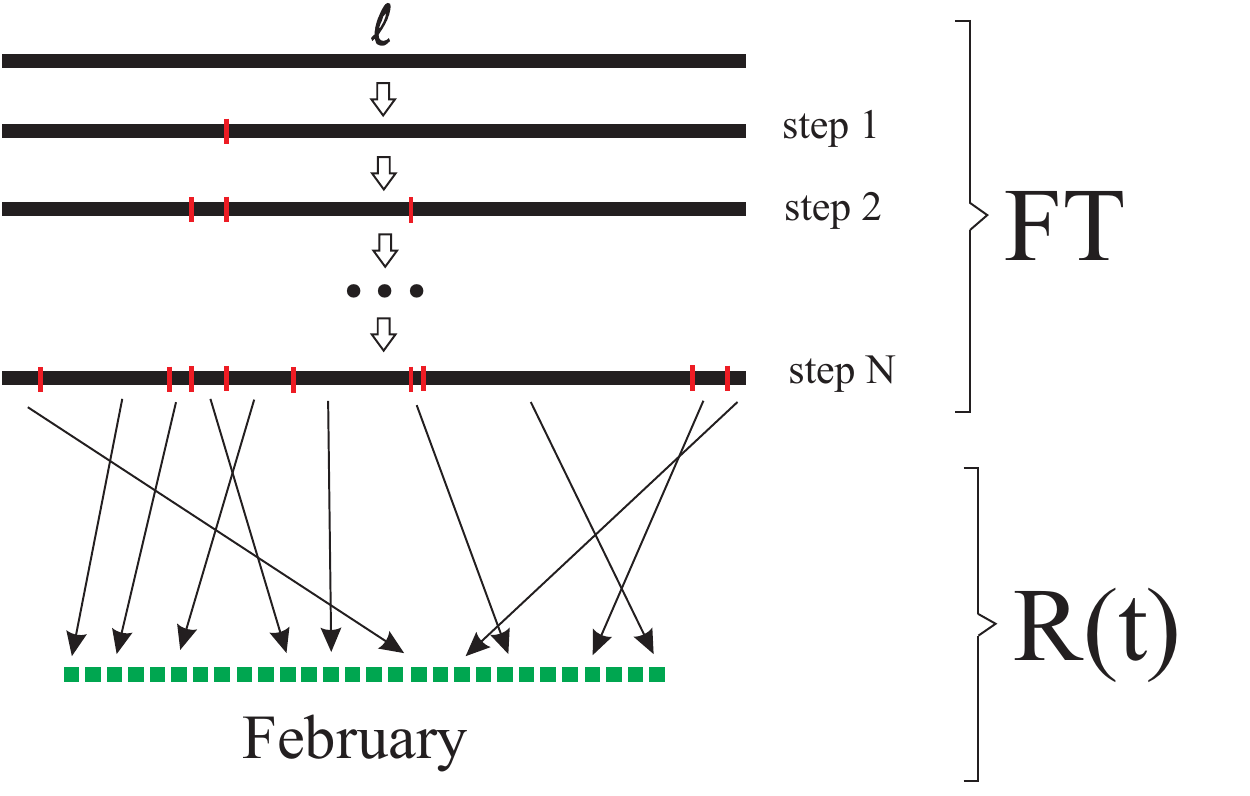}
    \put(90,56){}
  \end{overpic}\hspace{0.3cm}\vspace{0.0cm}
\caption{Schematic of the construction of $R(t)$ corresponding to
  February. At the top of the figure we show the FT process and at the
  bottom the construction of $R(t)$, in which each square represents a
  day in February.  The FT process stops when the required number of
  subintervals (given by Eq.~(\ref{eq.ArtD})) is reached. The length
  of each of these segments represents the rainfall in one day in
  February, which is randomly chosen. Those days that are not
  associated with any length of the FT process, have
  $R(t)=0$.}\label{fig.AppAlpha}
\end{figure}

In order to model the variability in the daily rainfall $R(t)$, we
apply a FT process for each month, in which,
\begin{itemize}
\item the length of the initial segment $\ell$ is given by
  Eq.~(\ref{eq.ArtP}), $i.e.$, the monthly precipitation,
\item the number $D$ of subintervals is given by Eq.~(\ref{eq.ArtD}), $i.e.$, the number of rainy days.
\end{itemize}
After we apply the fracturing process, the length of each resulting
interval $\widetilde{\ell}_i$ (with $i=1,\cdots,D$) represents the total amount of
water that falls in the day $d_i$, which we choose at random as it is
shown in the schematic of Fig.~\ref{fig.AppAlpha}, and then we set
$R(d_i)=\widetilde{\ell}_i$.

\begin{figure}[H]
\centering
\vspace{0.5cm}
  \begin{overpic}[scale=0.25]{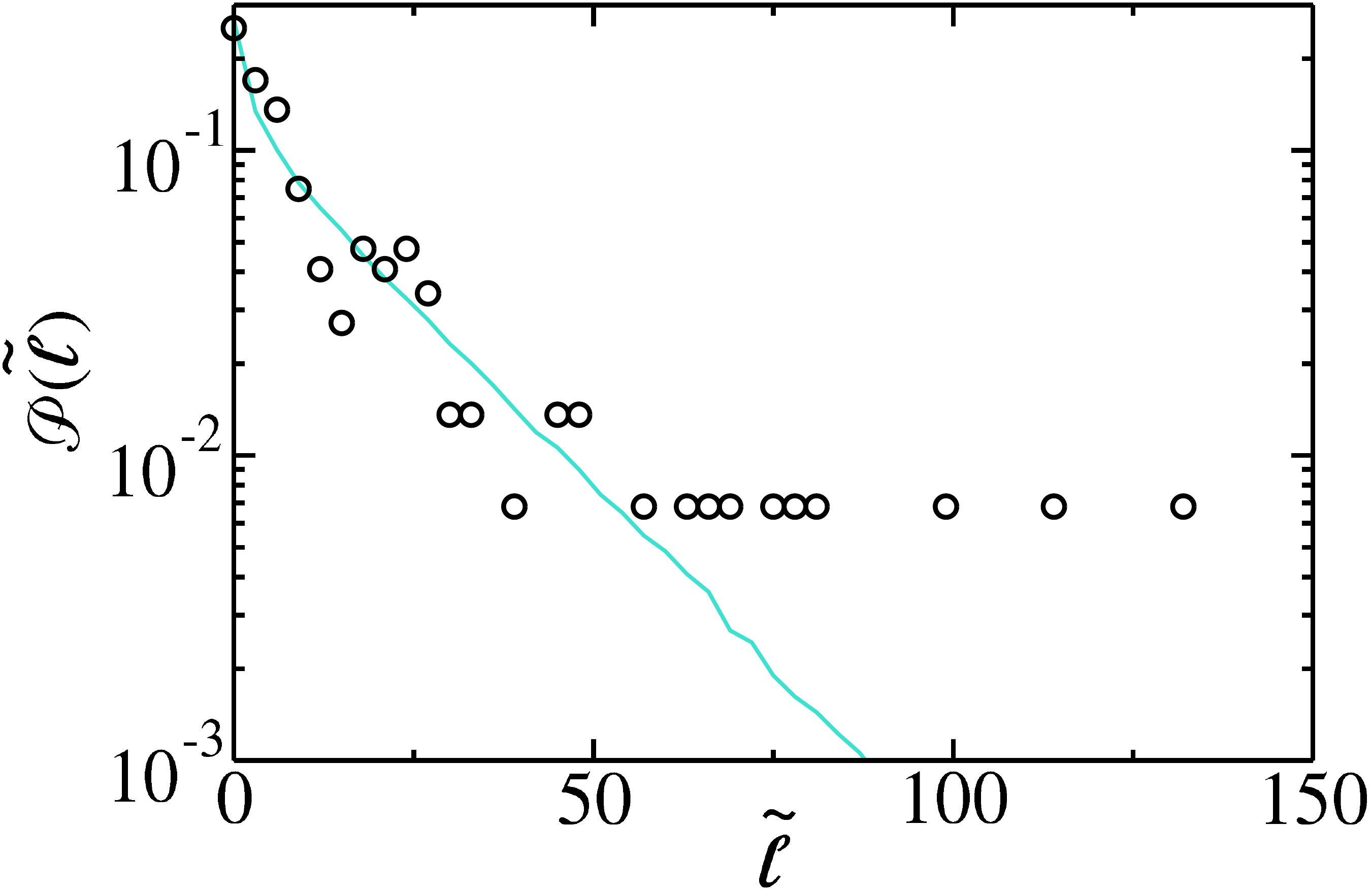}
    \put(90,56){}
  \end{overpic}\hspace{0.3cm}\vspace{0.0cm}
\caption{Observed February rainfall distribution in C\'ordoba during
  the period 2001-2015 (symbols) and distribution of segment lengths
  $\mathcal{P}(\widetilde{\ell})$ for $\ell=150$~mm, and $D=10$, and
  $\alpha=0.43$ (solid line) obtained over $2.10^4$
  realizations.}\label{fig.AlphaCord}
\end{figure}

In Fig.~\ref{fig.AlphaCord} we plot the distribution of segment
lengths obtained from an FT process for $\ell=150$~mm and $D=10$,
which are the average rainfall and the number of rainy days in
February, respectively. Here we use the value of $\alpha=0.43$ which
gives the best fit to the February rainfalls in the city of C\'ordoba in
the period 2001-2015.

\section{Sensitivity analysis}\label{Sec.Sensit}
A sensitivity analysis allows us to measure the impact of different
parameters on the relevant variables of our model. In this section, we
perform a one-way sensitivity analysis on the model of
Sec.~\ref{Sec.modd}, by varying a $\pm 25$\% of the baseline values of
individual parameters one at a time, while keeping the other
parameters constant in order to analyze their individual impact on the maximum
abundance of mosquitoes $M_{max}$. The parameters examined are: $\beta_L$,
$H_{max}$, $H_{min}$, and $K_{L}$. The results of the sensitivity analysis are
summarized in Table~\ref{tab.Sens}.

\begin{table}[H]
\centering
\caption{Variation of the maximum abundance of mosquitoes
  ($M_{max}=17.3$) when it is applied a one-way sensitivity
  analysis.}
\label{tab.Sens}
\begin{tabular}{|c|c|c|}
\hline
Parameter & -25\% & +25\% \\
\hline
$\beta_L$ &  -2.1\% & +1\%\\
$H_{max}$& -0.4\%  &+0.9\%\\
$H_{min}$& -0.7\% & +0.6\%\\
$K_L$ & -25\% & +25\% \\
\hline
\end{tabular}
\end{table}

It shows, as expected, that for higher values of $\beta_L$, $H_{max}$,
$H_{min}$, and $K_L$ the abundance $M_{max}$ increases. Although the
influence of the first three is rather weak, $M_{max}$ is heavily
influenced by $K_L$, which is therefore a critical parameter for the
estimation of mosquito abundance.

\section*{Acknowledgments}
This work was supported by SECyT-UNC (Projects 103/15 and 313/16),
CONICET (PIP 11220110100794), and PICT Cambio Clim\'atico (Ministerio
de Ciencia y T\'ecnica de la Provincia de C\'ordoba), PICT
Nro. 2013-1779 (ANPCYT-MYNCYT), Argentina. We also thank
Dr. A. M. Visintin and Bi\'ol. M. Beranek for useful discussions.
\section*{Bibliography}
\bibliographystyle{elsarticle-harv} 
\bibliography{bib}

\end{document}